\documentclass[usegraphicx]{mn2e}

\usepackage{graphicx,verbatim,amsmath,amssymb,amsfonts}
\usepackage{subfigure}
\usepackage{float}
\usepackage{color}

\begin{document}

\title[QSO UV LF evolution up to $z=8$]{Quasar UV luminosity function evolution up to $z=8$}
\author[S. Manti, S. Gallerani, A. Ferrara, B. Greig, C. Feruglio]
{S. Manti$^1$, S. Gallerani$^1$, A. Ferrara$^1$, B. Greig$^1$, C. Feruglio$^1$
\\
$^1$Scuola Normale Superiore, Piazza dei Cavalieri 7, 56126, Pisa, Italy}
\date{\today} 
\pagerange{\pageref{firstpage}--\pageref{lastpage}} \pubyear{2015} 
\maketitle

\setlength\parindent{12pt}

\begin{abstract}
We study the redshift evolution of the quasar UV Luminosity Function (LF) for $0.5<z < 6.5$, by collecting the most up to date observational data and, in particular, the recently discovered population of faint AGNs. We fit the QSO LF using either a double power-law or a Schechter function, finding that both forms provide good fits to the data. We derive empirical relations for the LF parameters as a function of redshift and, based on these results, predict the quasar UV LF at $z=8$. 
From the inferred LF evolution, we compute the redshift evolution of the QSO/AGN comoving ionizing emissivity and hydrogen photoionization rate. If faint AGNs are included, the contribution of quasars to reionization increases substantially. However, their level of contribution critically depends on the detailed shape of the QSO LF, which can be constrained by efficient searches of high-$z$ quasars. To this aim, we predict the expected (i) number of $z>6$ quasars detectable by ongoing and future NIR surveys (as EUCLID and WFIRST), and (ii) number counts for a single radio-recombination line observation with SKA-MID (FoV = 0.49 $\text{deg}^2$) as a function of the H$n\alpha$ flux density, at $0<z<8$. These surveys (even at $z<6$) will be fundamental to better constrain the role of quasars as reionization sources.
\end{abstract}


\section{Introduction}
The study of the quasar Luminosity Function (QSO LF) is of great relevance due to several reasons. First, it represents one of the most important observational probes of the growth of supermassive black holes (SMBHs) over cosmic time (e.g. Yu $\&$ Tremaine 2002; Li et al. 2007; Tanaka $\&$ Haiman 2009
). Indeed, many observations (Fan et al. 2006; Venemans et al. 2007, 2013; Jiang et al. 2008, 2009; Willott et al. 2010a; Morganson et al. 2012; Ba$\tilde{\text{n}}$ados et al. 2014) have shown that $z\sim 6$ quasars harbor SMBHs with masses $M_{\bullet} \sim (0.02-1.1)\times10^{10} M_{\odot}$ (Barth et al. 2003; Priddey et al. 2003; Willott et al. 2003, 2005; Jiang et al. 2007; Wang et al. 2010; Mortlock 2011; Wu et al. 2015), which have grown rapidly from their initial smaller ``seed'' black holes. The details of this growth are largely uncertain (Tanaka $\&$ Haiman 2009; Volonteri 2010; Treister 2013; Ferrara et al. 2014; Lupi et al. 2014; Tanaka 2014; Volonteri $\&$ Silk 2014), so the LF can provide vital additional inputs to theoretical models.

Second, the evidence of a strong redshift evolution in the quasar/AGN population (Schmidt 1968, 1972; Braccesi et al. 1980; Schmidt $\&$ Green 1983; Boyle et al. 1988, 2000; Goldschmidt $\&$ Miller 1998; Hewett et al. 1993; Croom et al. 2004; Richards et al. 2006), showing an increase in the number density of these sources with time up to $z\sim 2.5$ followed by a decline\footnote{Due to the peak in AGN activity, $z\sim 2-3$ has been often designated as the ``quasar epoch'', i.e. the period where QSOs were most active.} (Osmer 1982; Warren et al. 1994; Schmidt et al. 1995; Fan et al. 2001), has challenged our physical understanding so far. 

Following ``Pure Luminosity Evolution'' (PLE) models, the QSO LF is canonically described by four parameters: the normalization $\Phi^*$, the break magnitude $M^*$, the faint-end, $\alpha$, and bright-end, $\beta$, slopes.
In such representation, the overall normalization and the power-law slopes are independent of redshift (Boyle et al. 1988, 2000; Pei 1995). Nevertheless, many studies have shown that while PLE works well at $z<2.3$ (Boyle et al. 1988, 2000), it fails at higher redshifts, indicating an evolution of the QSO LF in large quasar samples (Schmidt et al. 1995; Fan et al. 2001; Richards et al. 2006; Hopkins et al. 2007; Croom et al. 2009).  However, the current picture is still rather sketchy. Croom et al. (2009) found a PLE form for the LF at $0.4<z<2.6$, with a decline in $\Phi^*$ and a steepening of $\beta$ from $z\sim 0.5$ ($\beta \sim -3.0$) to $z\sim 2.5$ ($\beta \sim -3.5$). Hunt et al. (2004) performed the first measurement of the faint-end at $z\sim 3$, finding a relatively shallow slope value $\alpha\sim -1.24$.  Using SDSS data in $z=0-5$, Richards et al. (2006) derived a redshift-independent slope $\beta=-3.1$ at $z<2.4$ and a flattening of $\beta$ above $z\sim 2.4$. Additional studies have attempted to obtain the LF at $z\sim 6$. Among these, Willott et al. (2010a) found a lower value of $\Phi^*\sim 10^{-8}$ Mpc$^{-3}$ mag$^{-1}$, and a flatter bright-end slope ($-3.8<\beta<-2.3$ for $-26<M^*<-24$) than derived in previous analyses (Fan et al. 2003, 2004; Richards et al. 2004; Jiang et al. 2008, 2009). Finally, Kashikawa et al. (2015) found a steeper faint-end slope than what is reported at lower redshifts.

In addition, observations with X-ray surveys (e.g. Hasinger et al. 2001; Giacconi et al. 2002; Alexander et al. 2003; Worsley et al. 2004) have shown that the space density of brighter sources peaks at higher redshifts than those of less luminous objects, a phenomenon known as ``cosmic downsizing'' of the quasar activity (Cowie et al. 2003; Ueda et al. 2003; Heckman et al. 2004; Merloni 2004; Barger et al. 2005; Hasinger et al. 2005; Croom et al. 2009).
Solidly assessing the redshift evolution of the LF \textit{shape} is then crucially important to clarify the physical mechanisms of black hole accretion/growth and AGN activity. For example, the bright end of the QSO LF provides information on quasar properties during Eddington-limited accretion phases (Hopkins et al. 2005; Willott et al. 2010b; June et al. 2015); the faint end is instead related to the duration of low accretion rate phases (Hopkins et al. 2007). 

In addition to QSO internal processes, an accurate determination of high-$z$ QSO LFs might enable us to set more stringent limits on the ionizing photon production by these sources during the Epoch of Reionization (EoR). This would be a major result as at present the relative contribution from stars in galaxies and accreting sources such as quasars/AGNs is far cry from being known.

The usual argument against QSOs being important is based on their sharply decreasing number density at $z>3$, which prevents AGNs from producing enough ionizing emissivity at $z>4$ (Masters et al. 2012). Therefore, the high-$z$ population of star-forming galaxies is thought to be the most natural candidate for cosmic reionization (e.g. Madau 1991; Haardt $\&$ Madau 1996, 2012; Giallongo et al. 1997; Willott et al. 2010a; Bouwens et al. 2012), provided that at least $>20$\% of the ionizing photons escape into the IGM, a non-trivial requirement.  


However, such persuasion has been shaken by the recent results from Giallongo et al. (2015), who found evidence for a new population of faint quasars ($-22.5\lesssim M_{\text{AB}}^{1450}\lesssim -18.5$) at $4<z<6.5$. 
Based on these observations, Madau $\&$ Haardt (2015) (hereafter MH15) obtained an upward revision of the comoving AGN emissivity ($\epsilon_Q=2.5\times10^{24}$ erg s$^{-1}$ Mpc$^{-3}$ Hz$^{-1}$), of $\sim 10$ times with respect to Hopkins et al. (2007) (H07; see also Haardt $\&$ Madau 2012 (HM12)), and potentially sufficient to keep the IGM ionized at $z=6$. If these claims are confirmed, the contribution of quasars to cosmic reionization could be more relevant than previously estimated (e.g. Glikman et al. 2011). However, it is worth noting that the impact of AGN radiation to the cosmic reionization is still controversial (e.g. Georgakakis et al. 2015; Kim et al. 2015) and requires further searches for faint quasars. 

In fact, given the lack of deep AGN surveys at various wavelengths (Shankar $\&$ Mathur 2007), only a very small number of low-luminosity quasars ($M_{\text{AB}}^{1450} > -24$) at $z\gtrsim 6$ has been spectroscopically identified (Willott et al. 2009; Kashikawa et al. 2015; Kim et al. 2015). This implies that both the faint end of the $z\sim 6$ LF and the AGN role in reionization remain highly uncertain.  
%
\begin{figure*}
\centering   
 \begin{tabular}{@{}cc@{}}
\hspace{-1cm}\subfigure{\includegraphics[width=0.32\textwidth]{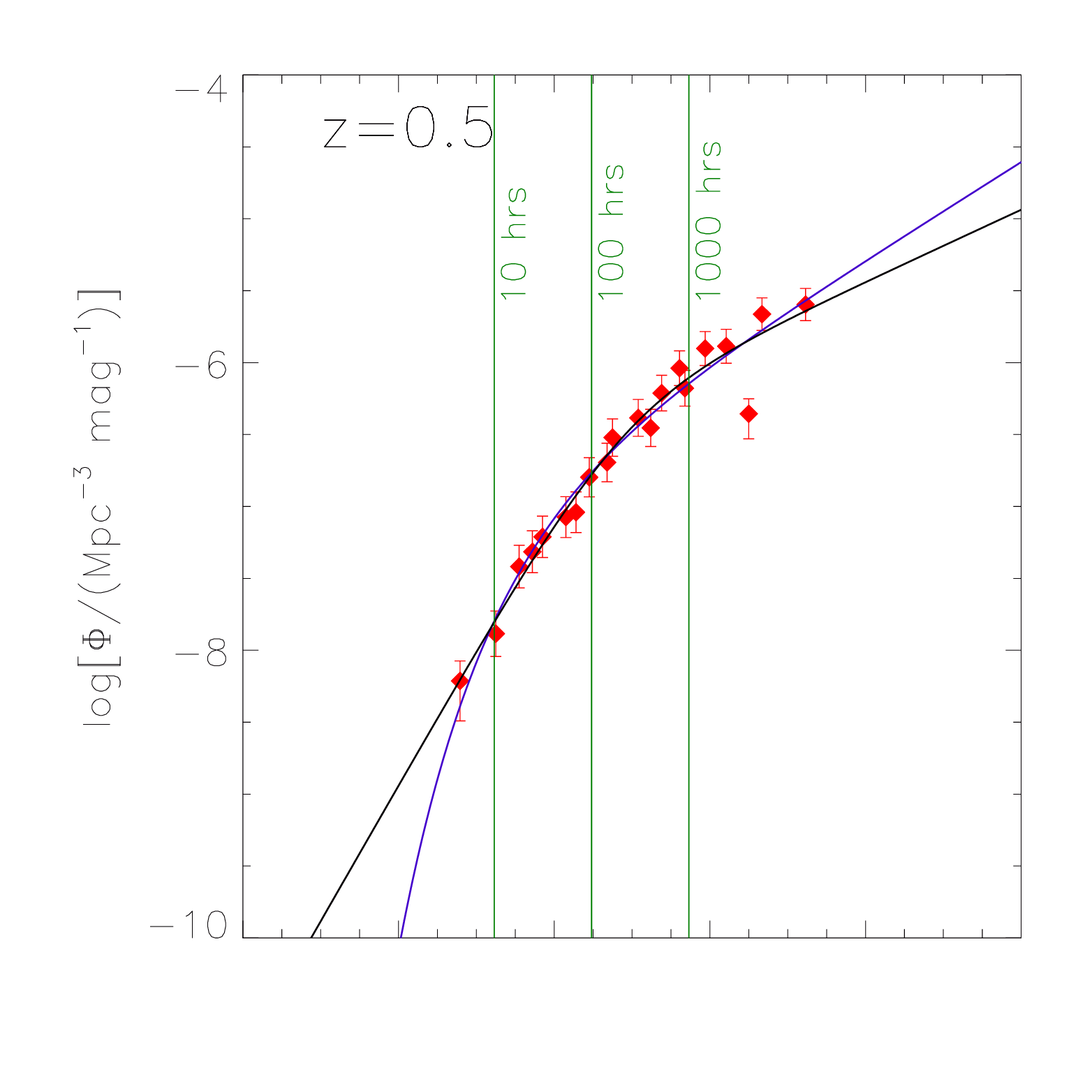}}
\hspace{-1.3cm}\subfigure{\includegraphics[width=0.32\textwidth]{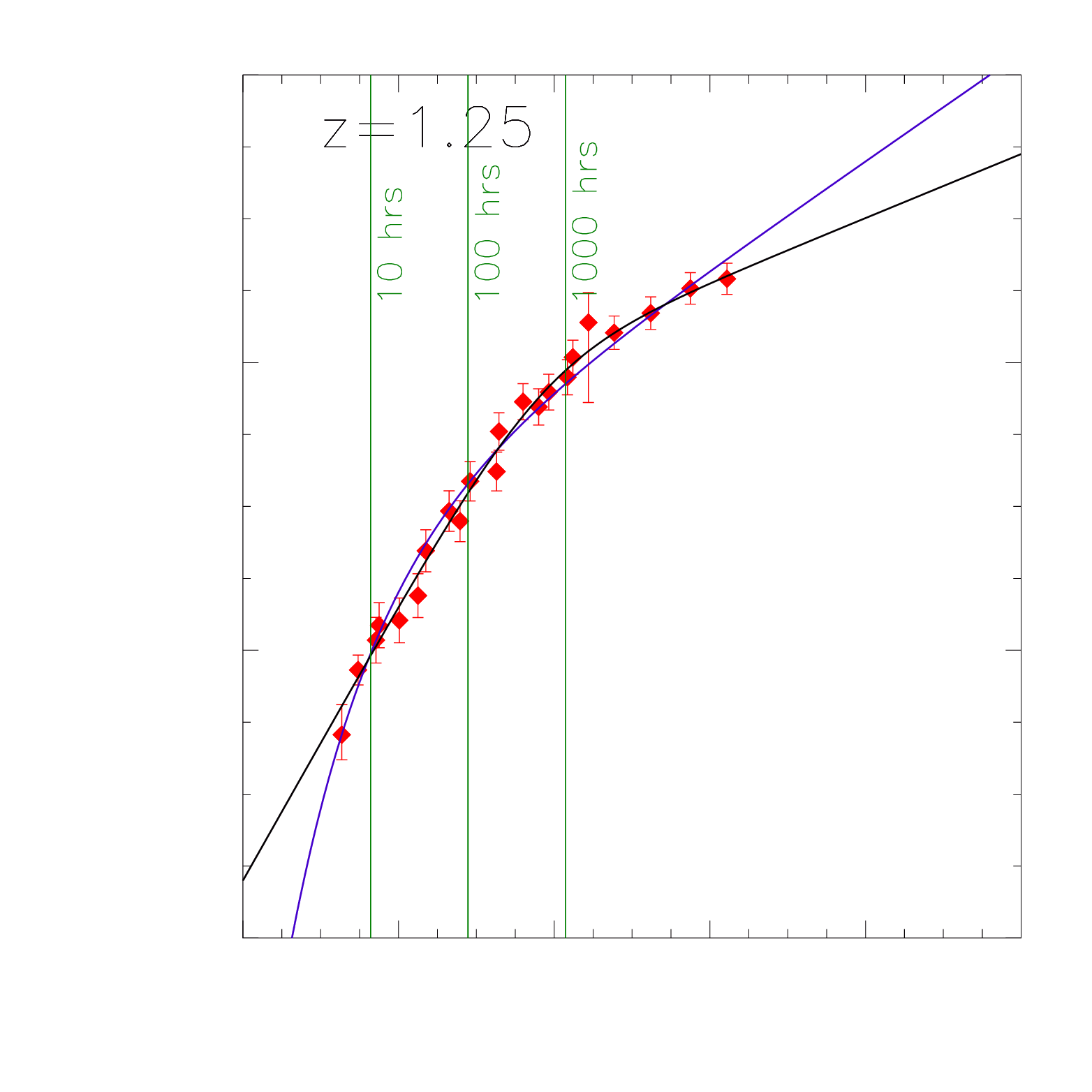}}
\hspace{-1.3cm}\subfigure{\includegraphics[width=0.32\textwidth]{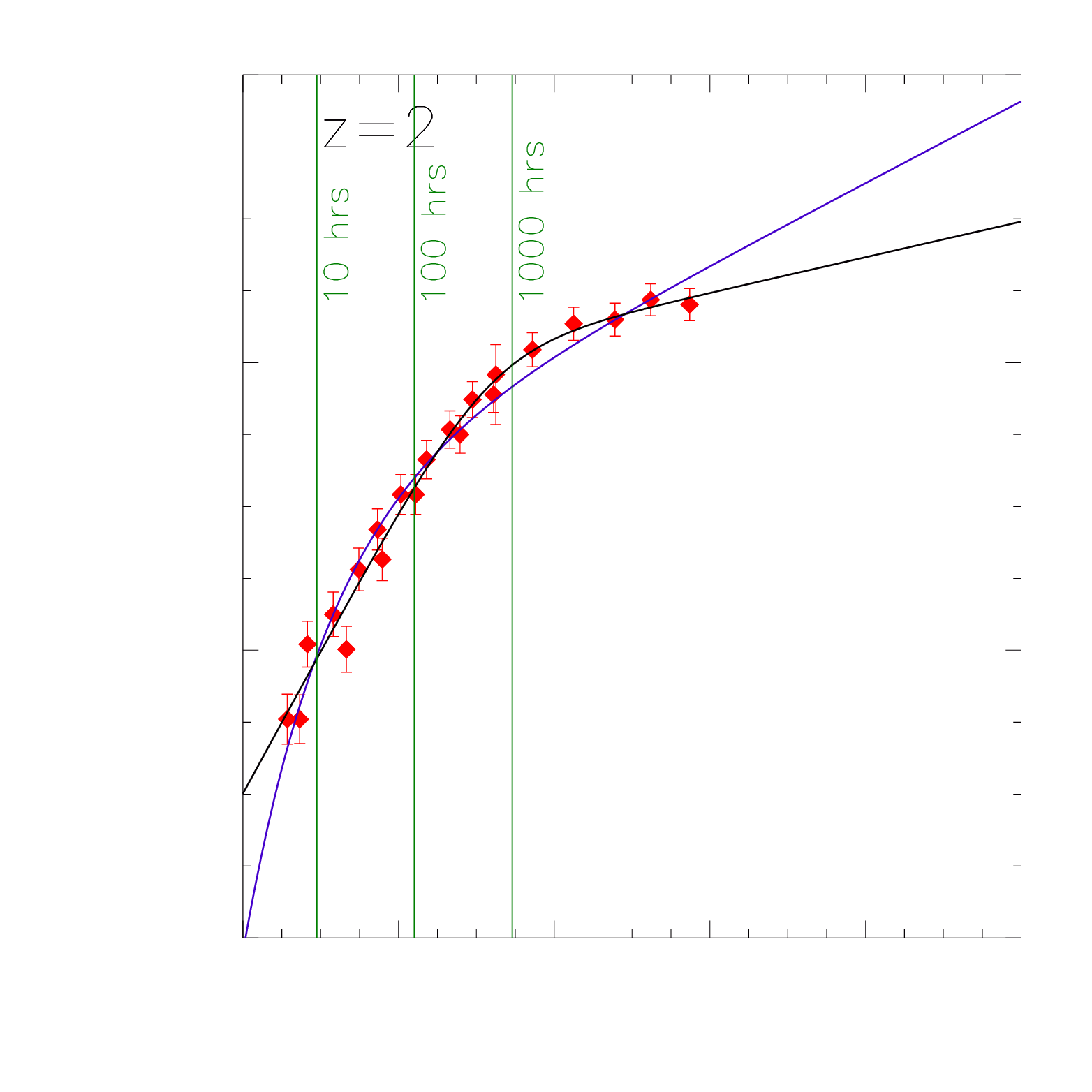}}
\hspace{-1.3cm}\subfigure{\includegraphics[width=0.32\textwidth]{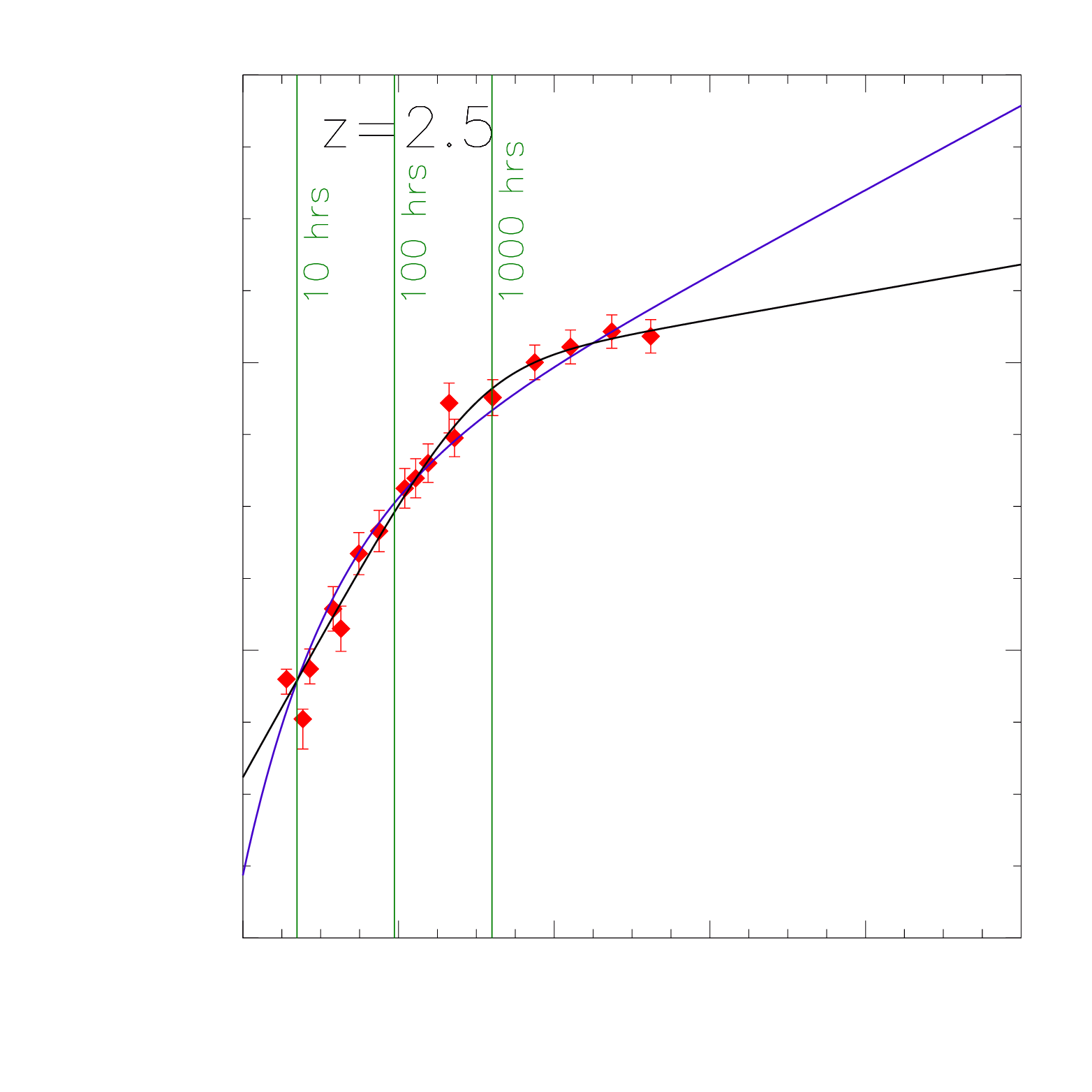}}
\vspace{-1.1cm}\\         
\hspace{-1.1cm}\subfigure{\includegraphics[width=0.32\textwidth]{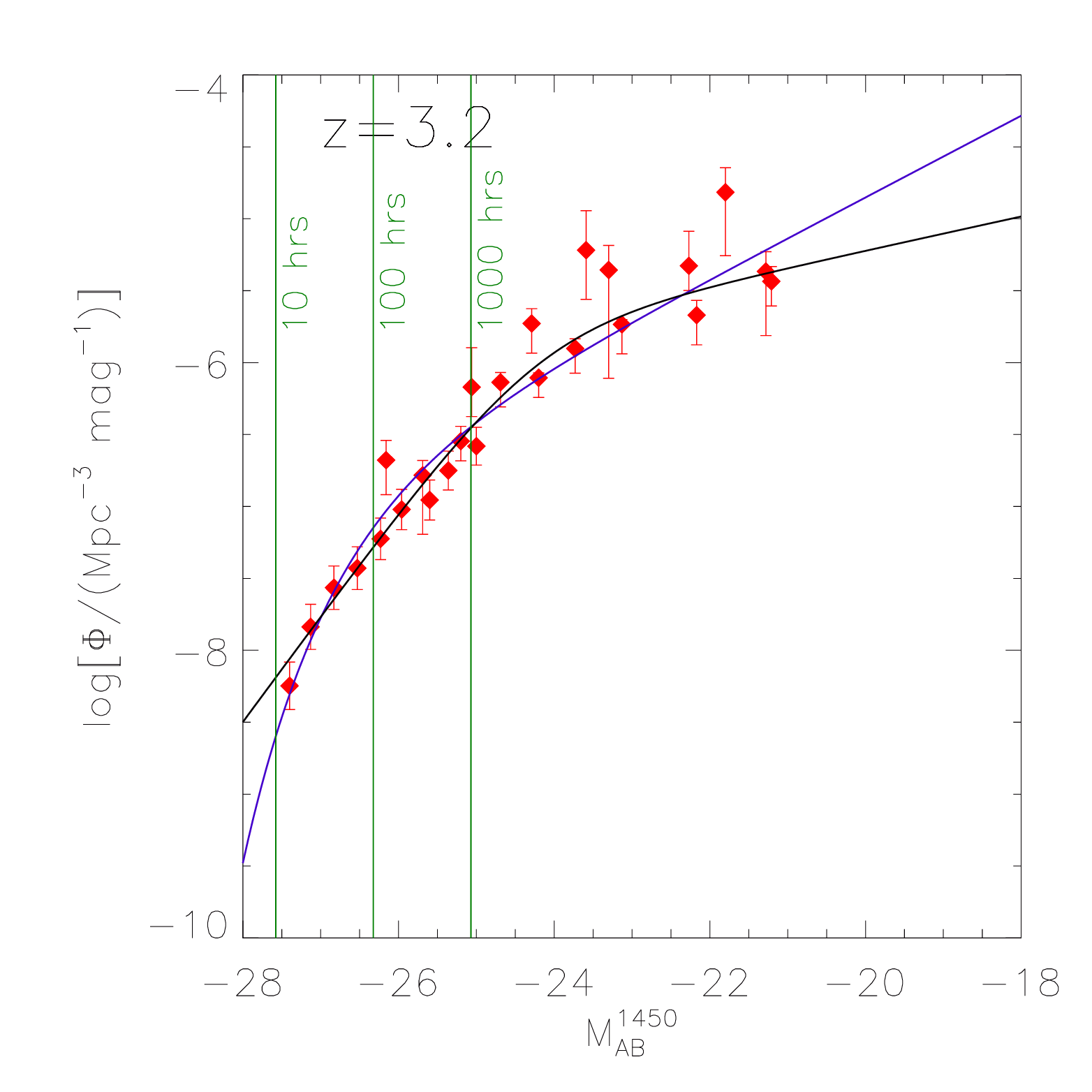}}
\hspace{-1.3cm}\subfigure{\includegraphics[width=0.32\textwidth]{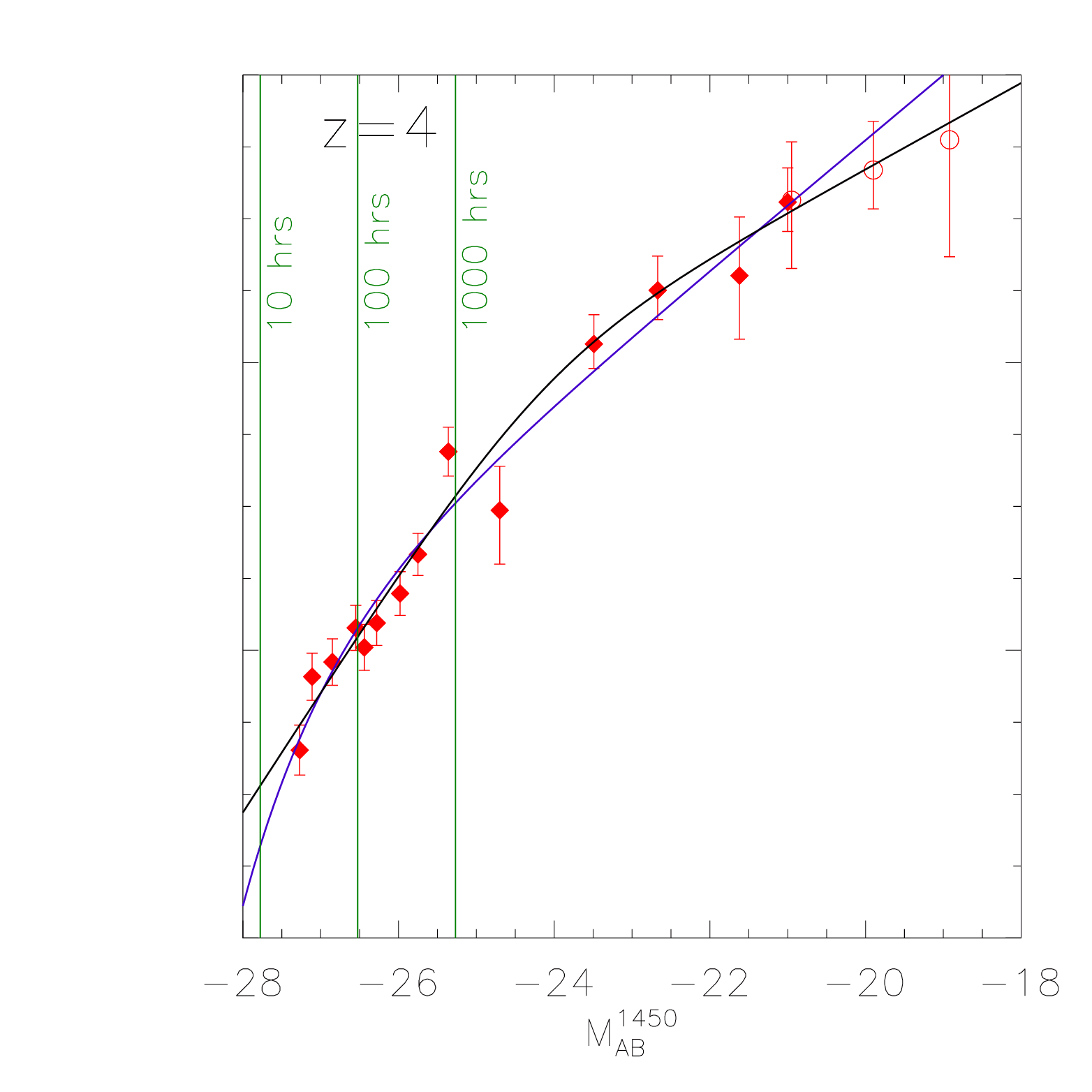}}
\hspace{-1.3cm}\subfigure{\includegraphics[width=0.32\textwidth]{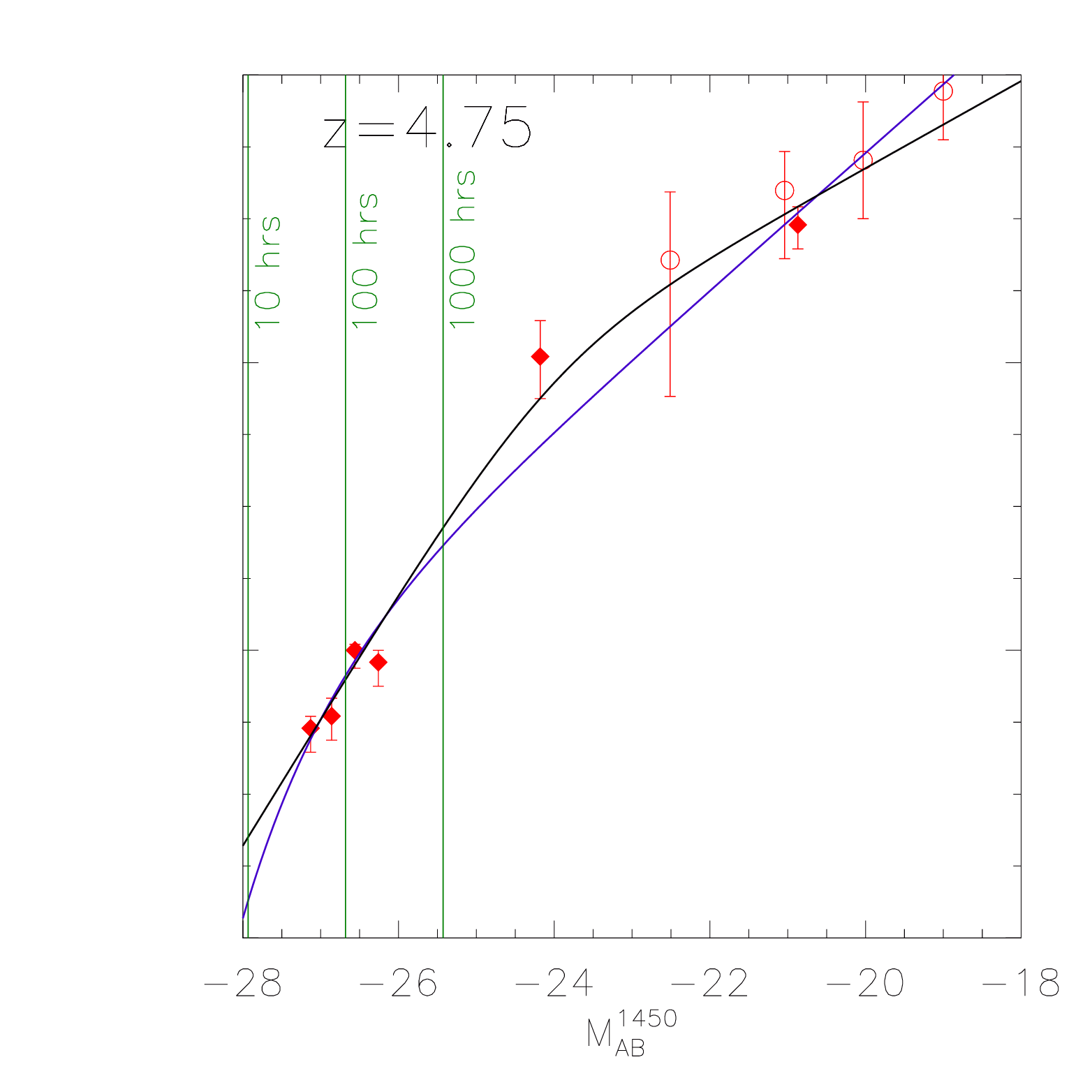}}
\hspace{-1.3cm}\subfigure{\includegraphics[width=0.32\textwidth]{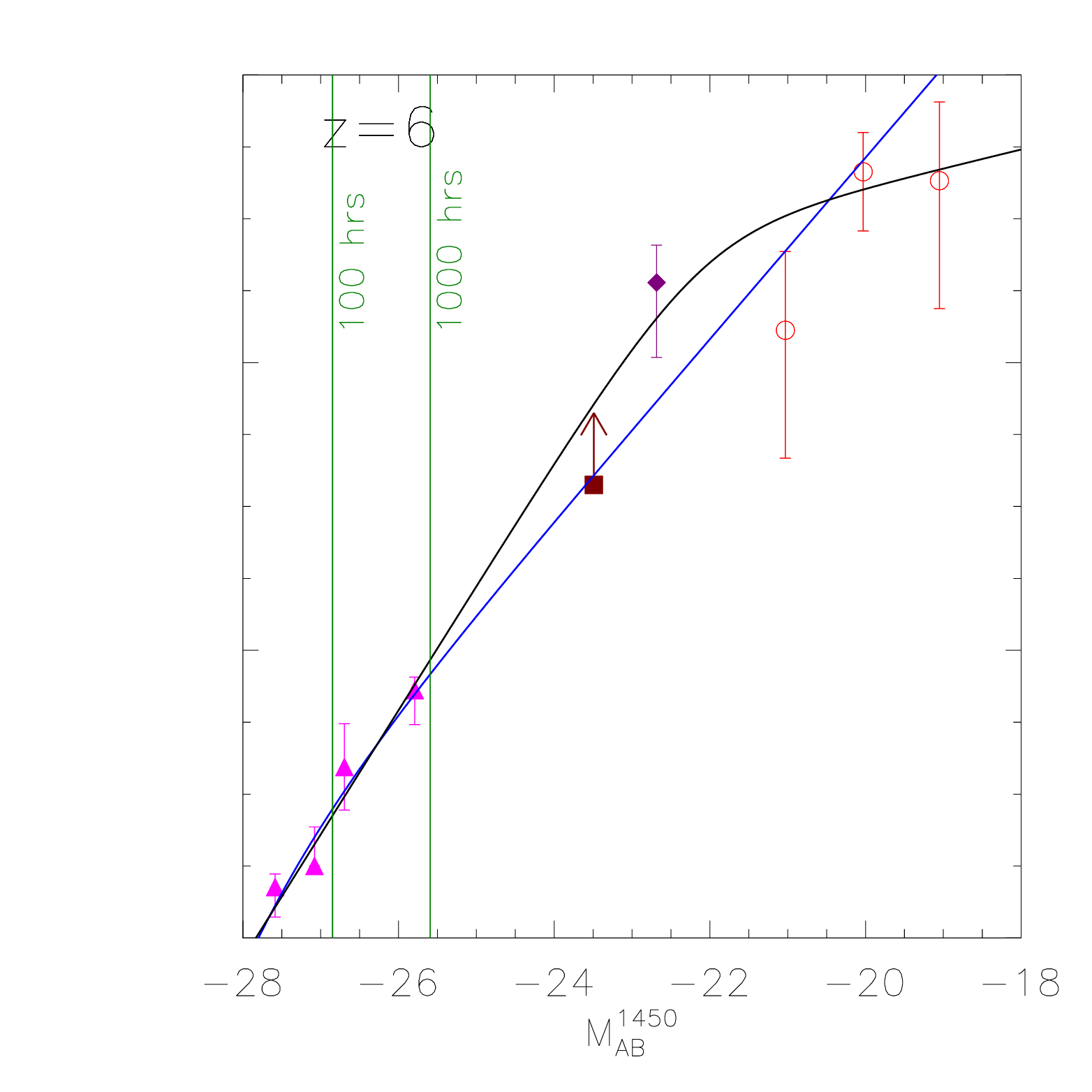}}
\end{tabular}  
\caption{Evolution of the UV luminosity function $\log\Phi$ $(\text{Mpc}^{-3}\hspace{0.1cm}\text{mag}^{-1})$ as a function of the absolute magnitude $M_{\text{AB}}^{1450}$, from $z=0.5$ to $z=6$. In each panel, symbols represent observational data with the corresponding error bars (Jiang et al. 2009: magenta triangles; Fiore et al. 2012: purple diamond; Giallongo et al. 2012 -- and references therein --, Giallongo et al. 2015: red circles, filled and empty respectively; Kashikawa et al. 2015: dark red square), while the solid curves represent our best-fit parameterization of the QSO LF using a double power-law function (black) and a Schechter function (blue). In each panel, the green vertical lines indicate the 5$\sigma$ detection thresholds with SKA-MID at $\nu=13.8$ GHz, in 10, 100 and 1000 hours of observing time.
}
   \label{lumFunc}
\end{figure*}
Optical/NIR surveys are limited by the fact that high-$z$ quasars are likely to be heavily obscured during their growth phase. In this case one can exploit obscuration-insensitive experiments, such as radio recombination line emission (see e.g.\ Manti et al. 2016, hereafter M16). This technique will become possible thanks to the extraordinary capabilities of the SKA (Square Kilometre Array) telescope in terms of frequency coverage ($0.35-13.8$ GHz), angular resolution and sensitivity (Morganti et al. 2014), and might even allow the detection of Compton-thick sources. 

Motivated by the above issues, based on a compilation of the most updated observational data at different redshifts ($z=0.5$ to $z=6.5$; Jiang et al. 2009; Fiore et al. 2012; Giallongo  et al. 2012, 2015; Kashikawa et al. 2015), we provide a comprehensive overview of the evolution of the quasar UV LF at $0<z<6.5$ in terms of its parameters. These results are then used to predict\footnote{Throughout this paper we assume a flat cosmological model with $H_0=67.3$ km s$^{-1}$ Mpc$^{-1}$, $\Omega_m=0.315$ and $\Omega_\Lambda=0.685$.} the expected number densities and number counts of AGNs up to  $z=8$.

\section{QSO LF evolution}
In order to study the evolution of the quasar UV LF, we first collect the most up to date measurements from $z=0.5$ to $z=6.5$:
\begin{itemize}
\item $0.5<z<6$ by Giallongo et al. (2012), and references therein (Croom et al. 2009; Richards et al. 2009 at $z\leq 2.5$, Bongiorno et al. 2007; Fontanot et al. 2007; Siana et al. 2008; Brusa et al. 2010; Civano et al. 2011; Glikman et al. 2011; Fiore et al. 2012 at $z>3$). For details on the number of quasars in each sample, the instrument used for the detection, the magnitude limit of the survey, and the minimum and maximum $M_{\text{AB}}^{1450}$ magnitude of each QSO sample see the corresponding reference.
\item $4<z<6.5$ by Giallongo et al. (2015). The 22 AGN candidates of this sample have been selected in the CANDELS GOODS-South field, and are characterized by $-22.5\lesssim M_{\text{AB}}^{1450}\lesssim -18.5$. The selection has been done in the NIR $H$ band down to very faint levels ($H\leq 27$) using reliable photometric redshifts.
\item $5.78<z<6.00$ by Jiang et al. (2009). The sample is composed of 4 new quasars among six which have been selected from the Sloan Digital Sky Survey (SDSS) southern survey. It constitutes a flux-limited sample with $21 < z_{\text{AB}} < 21.8$ over an effective area of $195$ deg$^2$, and is 
 combined with 17 QSOs at $z>5.7$ selected in the SDSS main survey using similar criteria (these form a flux-limited sample with $z_{\text{AB}} < 20$ over $\sim 8000$ deg$^2$) and a flux-limited sample of 6 quasars with $z_{\text{AB}}<21$ (Jiang et al. 2008).
\item $z>5.8$ by Fiore et al. (2012). 30 new AGN candidates have been identified at $z>3$ in the \textit{Chandra} deep field south (CDFS) using the 4 Ms \textit{Chandra} observation. Among them, the $z>5.8$ candidates are characterized by $43.5<\log\text{L}(2-10\text{ keV})<44.5$ (the luminosity limit at $z=7.5$ is $\log\text{L}(2-10\text{ keV})\sim 43.3$).
\item $6.041<z<6.156$ by Kashikawa et al. (2015). The sample includes 2 AGN candidates discovered with the Subaru/Suprime-Cam, and identified through follow-up spectroscopy (one apparent quasar with $M_{\text{AB}}^{1450}=-23.10$ at $z=6.156$ and one possible quasar with $M_{\text{AB}}^{1450}=-22.58$ at $z=6.041$).
\end{itemize}

\begin{table*}
\begin{minipage}{115mm}
\caption{Best-fit values and corresponding errors of the QSO LF parameters at different redshifts (0.5 to 6), calculated via $\chi^2$ minimisation. Columns are: redshift, $\log[\Phi^*/(\text{Mpc}^{-3}\hspace{0.1cm}\text{mag}^{-1})]$, $M^*$, $\alpha$, $\beta$, $\gamma$, and the $\chi^2$ of the fit at each $z$. For each $z$, the first (second) row refers to a double power-law (Schechter) fit.}
\centering
\begin{tabular}{@{}cccccc||c}
\hline
Redshift & $\log(\Phi ^{*})$ & $M^{*}$ & $\alpha$ & $\beta$ & $\gamma$ & $\chi^2$\\
\hline\hline
$0.5$ & $-6.22^{+0.13}_{-0.13}$ & $-23.13^{+0.21}_{-0.22}$ & $-1.63^{+0.34}_{-0.35}$ & $-3.36^{+0.58}_{-0.55}$ & & $21.09$\vspace{0.1cm}\\
      & $-6.61^{+0.13}_{-0.13}$ & $-23.99^{+0.19}_{-0.22}$ &                         & & $-1.85^{+0.19}_{-0.19}$ & $23.39$\\\hline
$1.25$ & $-5.92^{+0.13}_{-0.13}$ & $-24.16^{+0.18}_{-0.19}$ & $-1.56^{+0.37}_{-0.37}$ & $-3.40^{+0.30}_{-0.30}$ & & $8.20$\vspace{0.1cm}\\
       & $-6.59^{+0.13}_{-0.13}$ & $-25.41^{+0.16}_{-0.16}$ &                        & & $-1.94^{+0.19}_{-0.18}$  & $15.71$\\\hline
$2$ & $-5.87^{+0.13}_{-0.13}$ & $-24.85^{+0.18}_{-0.19}$ & $-1.31^{+0.33}_{-0.33}$ & $-3.48^{+0.31}_{-0.30}$ & & $12.47$\vspace{0.1cm}\\
    & $-6.38^{+0.13}_{-0.13}$ & $-25.87^{+0.15}_{-0.18}$ &                         & & $-1.71^{+0.19}_{-0.19}$ & $24.46$\\\hline
$2.5$ & $-5.99^{+0.12}_{-0.12}$ & $-25.03^{+0.16}_{-0.17}$ & $-1.24^{+0.37}_{-0.37}$ & $-3.43^{+0.24}_{-0.23}$ & & $13.87$\vspace{0.1cm}\\
      & $-6.56^{+0.12}_{-0.12}$ & $-26.13^{+0.13}_{-0.15}$ &                       & & $-1.73^{+0.19}_{-0.19}$   & $27.15$\\\hline
$3.2$ & $-5.73^{+0.14}_{-0.14}$ & $-24.24^{+0.26}_{-0.28}$ & $-1.30^{+0.39}_{-0.40}$ & $-2.84^{+0.29}_{-0.28}$ & & $31.14$\vspace{0.1cm}\\
      & $-6.56^{+0.14}_{-0.14}$ & $-26.16^{+0.22}_{-0.26}$ &                       & & $-1.71^{+0.17}_{-0.17}$   & $38.53$\\\hline
$4$ & $-5.90^{+0.19}_{-0.19}$ & $-24.16^{+0.25}_{-0.26}$ & $-1.74^{+0.41}_{-0.42}$ & $-3.10^{+0.25}_{-0.24}$ & & $12.75$\vspace{0.1cm}\\
    & $-7.37^{+0.19}_{-0.19}$ & $-26.52^{+0.22}_{-0.23}$ &                       & & $-2.13^{+0.21}_{-0.21}$   & $16.95$\\\hline
$4.75$ & $-5.87^{+0.15}_{-0.15}$ & $-24.06^{+0.18}_{-0.18}$ & $-1.76^{+0.49}_{-0.24}$ & $-3.21^{+0.17}_{-0.16}$ & & $8.54$\vspace{0.1cm}\\
       & $-7.68^{+0.15}_{-0.15}$ & $-26.65^{+0.16}_{-0.18}$ &                       & & $-2.19^{+0.14}_{-0.14}$   & $12.25$\\\hline
$6$ & $-5.06^{+0.20}_{-0.20}$ & $-22.11^{+0.25}_{-0.24}$ & $-1.33^{+0.88}_{-0.93}$ & $-3.16^{+0.11}_{-0.11}$ & & $4.41$\vspace{0.1cm}\\
    & $-9.34^{+0.20}_{-0.20}$ & $-27.62^{+0.22}_{-0.25}$ &                       & & $-2.58^{+0.20}_{-0.20}$   & $9.51$\\
\hline\hline
\end{tabular}
\end{minipage}
\label{BF_param}
\end{table*}
\begin{figure*}
\centering      
\begin{tabular}{@{}cc@{}}
\hspace{-0.8cm}\subfigure{\includegraphics[width=0.28\textwidth]{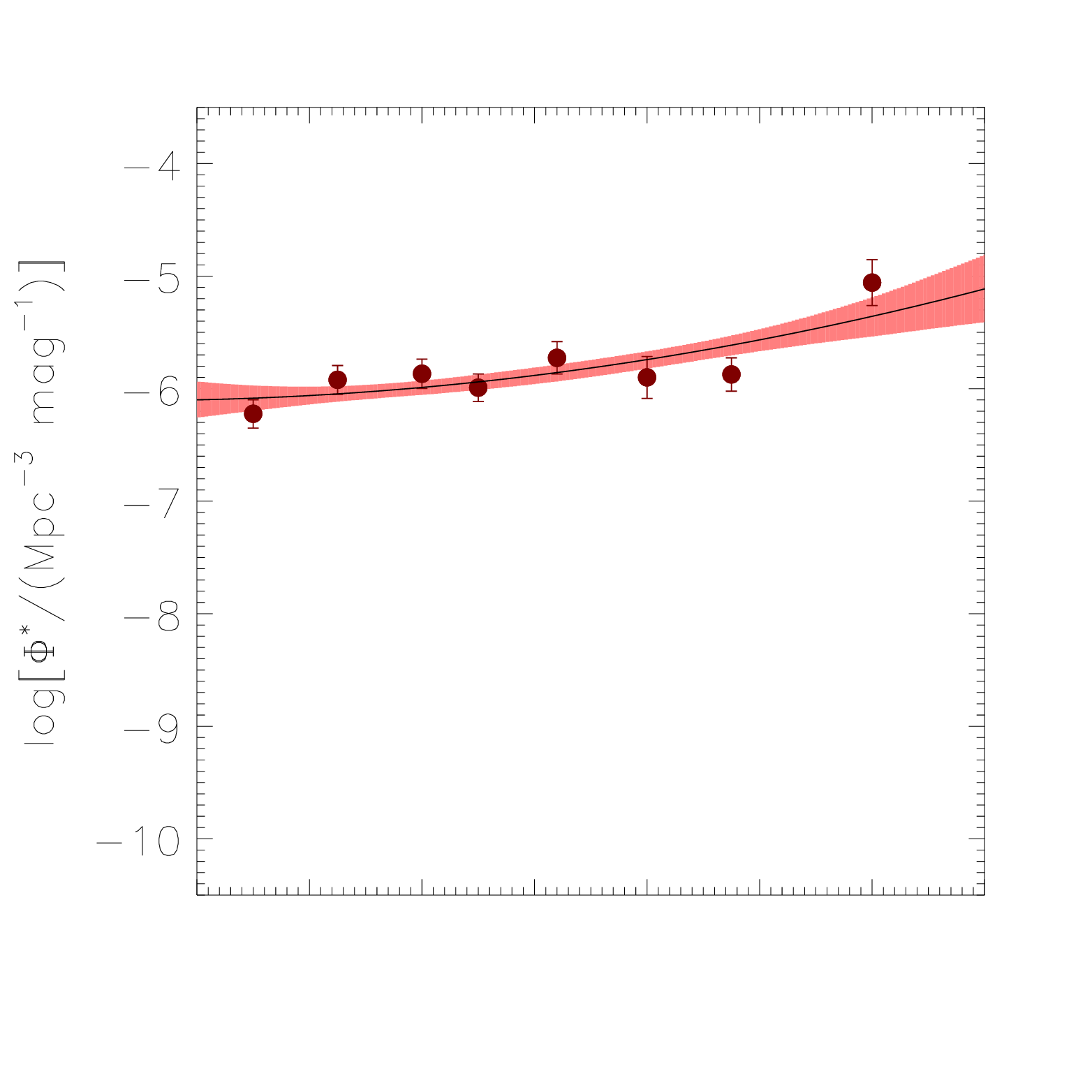}} 
\hspace{-0.4cm}\subfigure{\includegraphics[width=0.28\textwidth]{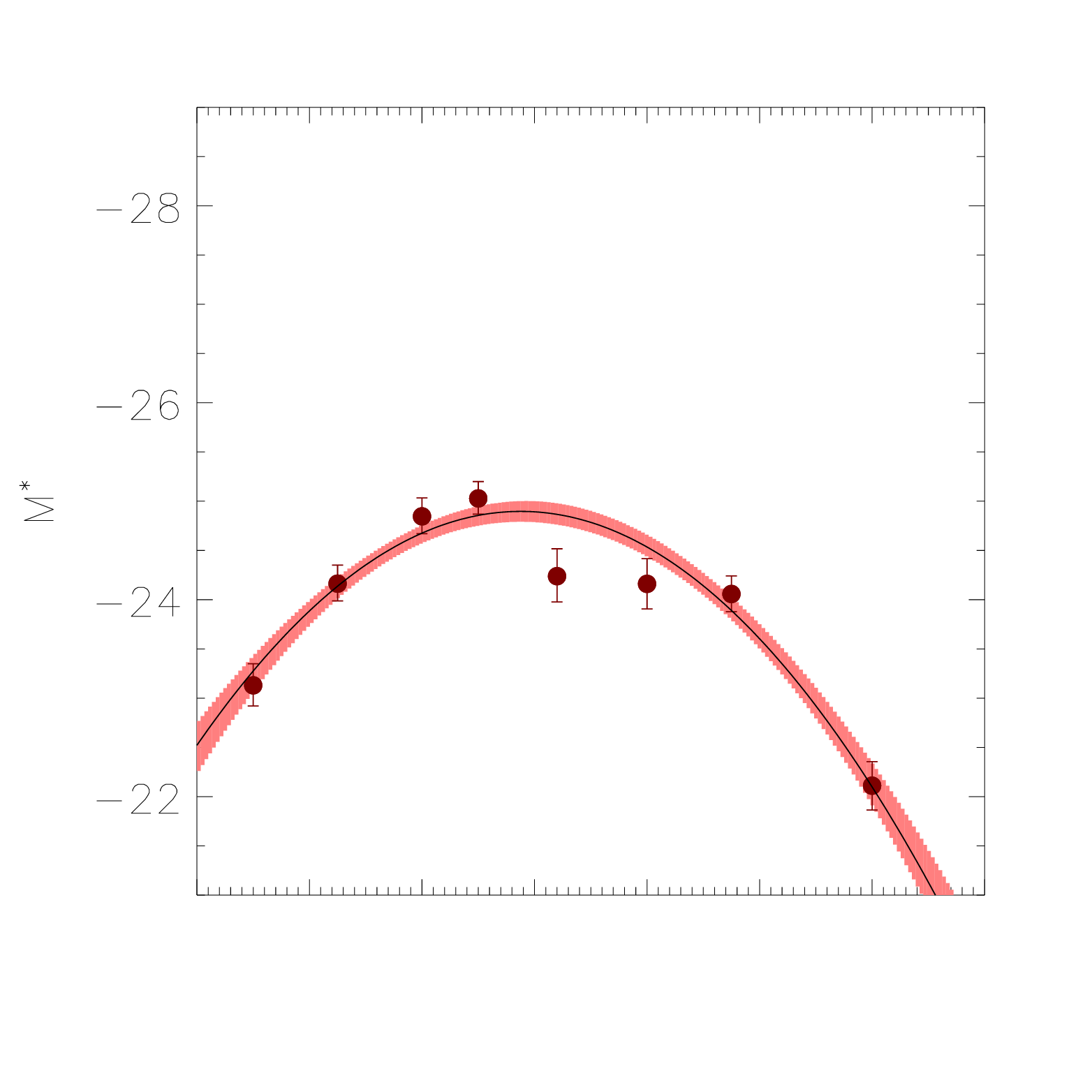}} 
\hspace{-0.4cm}\subfigure{\includegraphics[width=0.28\textwidth]{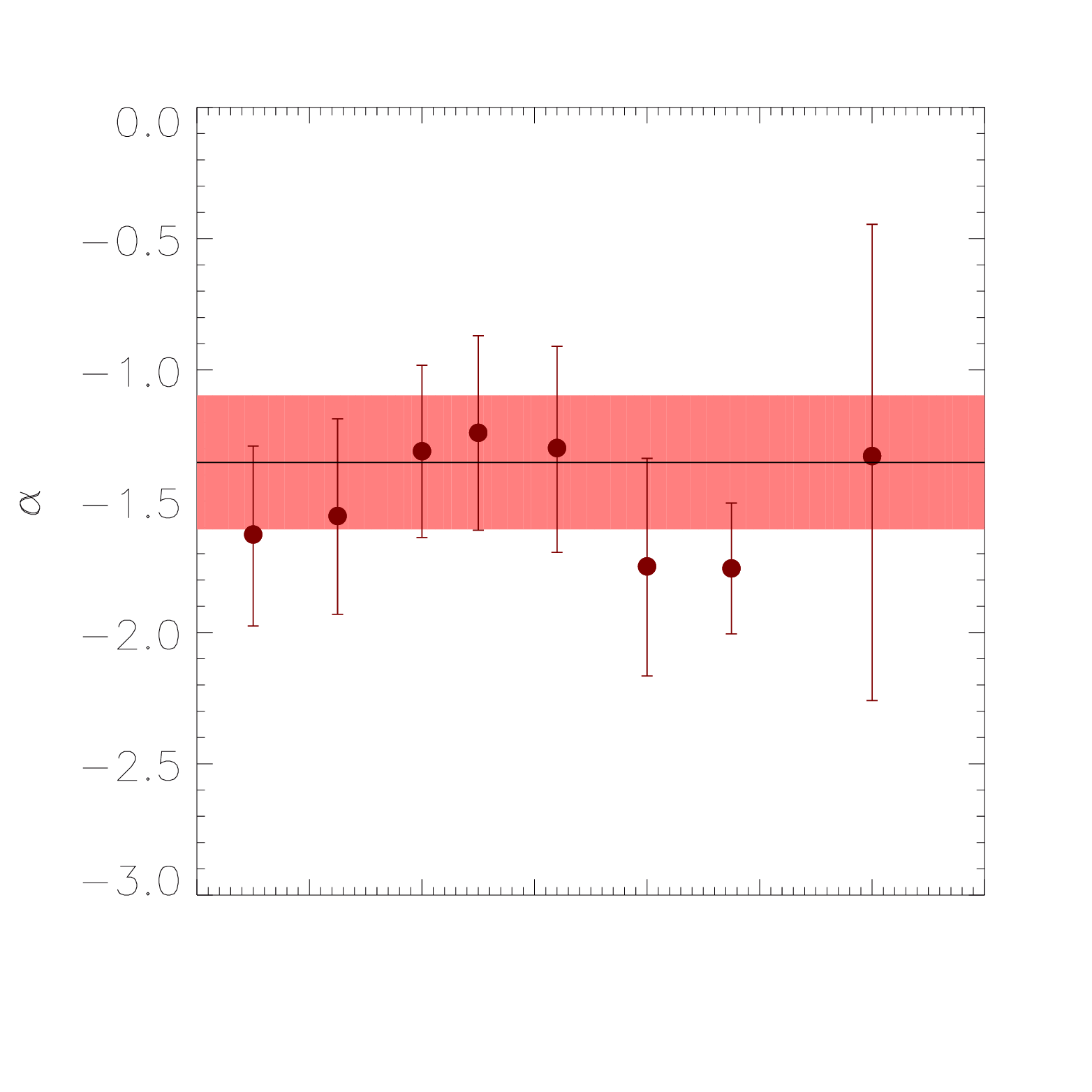}}
\hspace{-0.4cm}\subfigure{\includegraphics[width=0.28\textwidth]{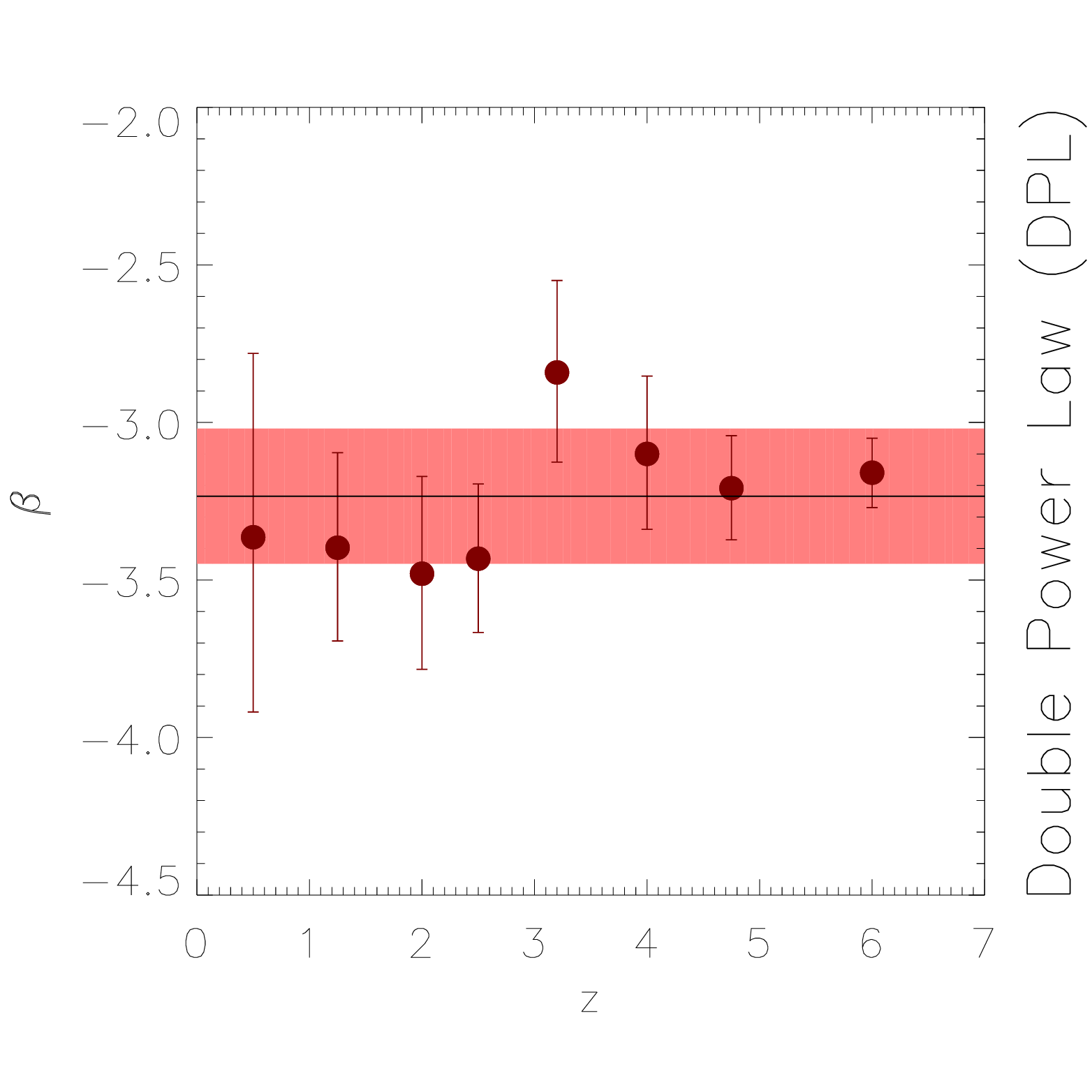}}
\vspace{-1.3cm}\\
\hspace{-5.6cm}\subfigure{\includegraphics[width=0.28\textwidth]{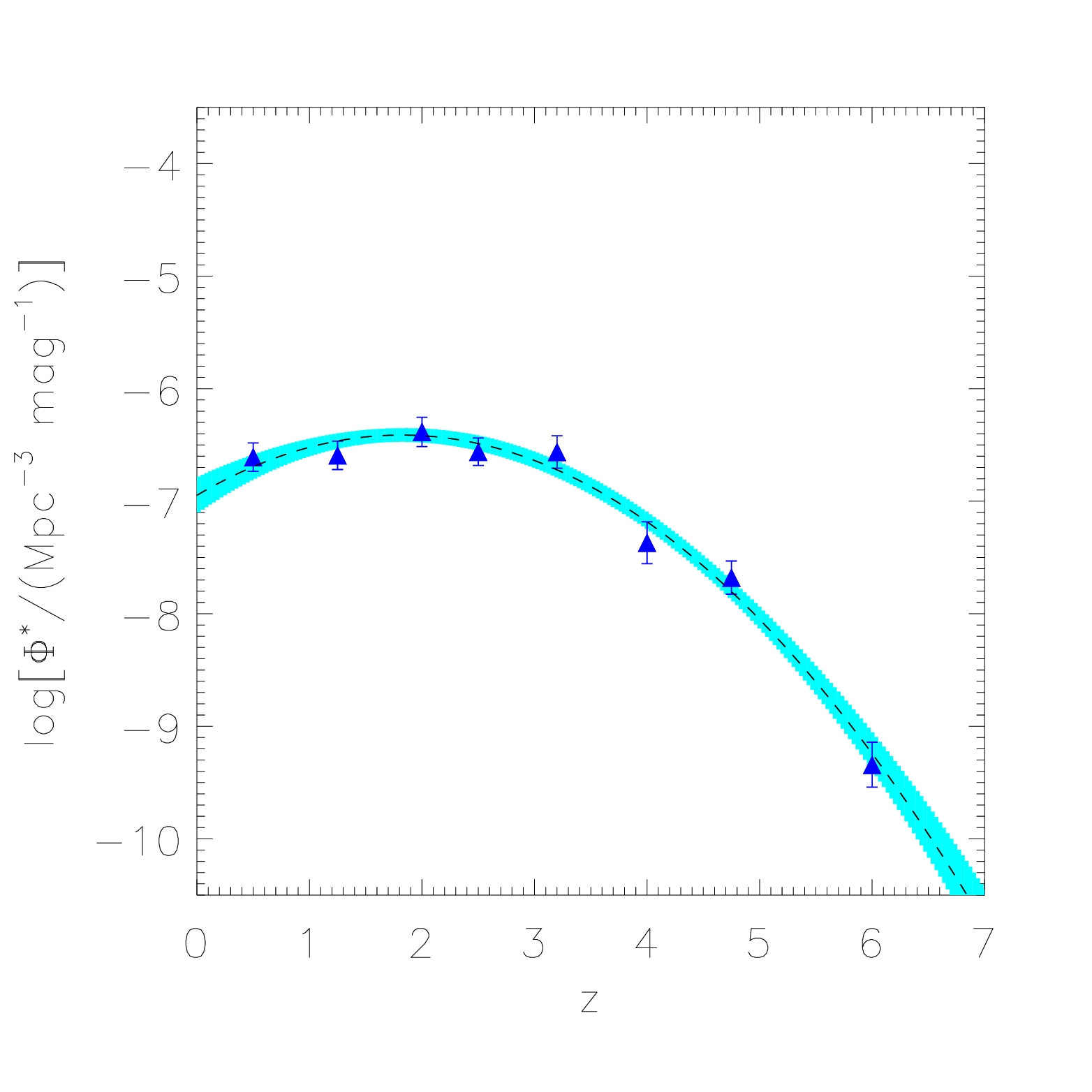}} 
\hspace{-0.4cm}\subfigure{\includegraphics[width=0.28\textwidth]{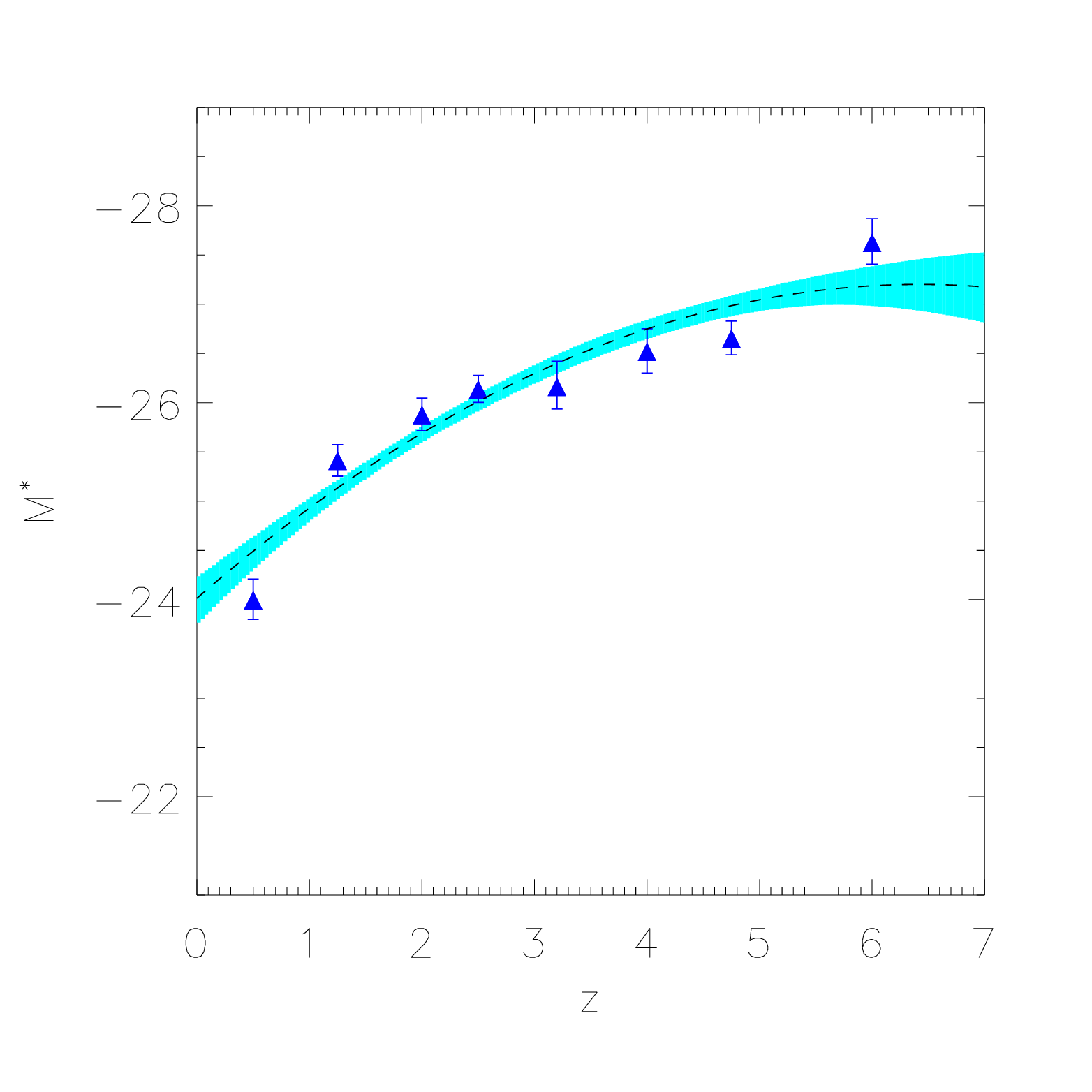}} 
\hspace{-0.4cm}\subfigure{\includegraphics[width=0.28\textwidth]{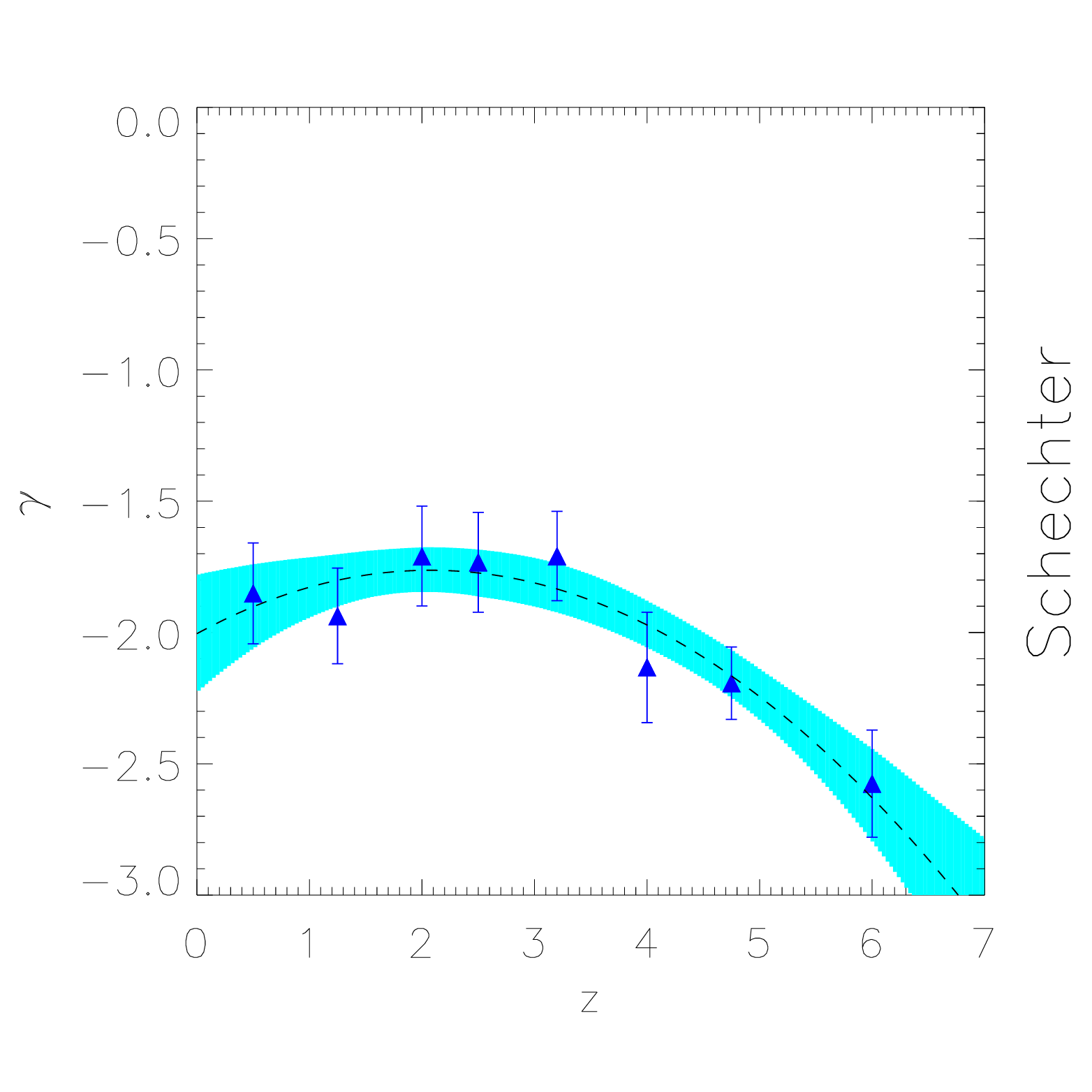}} 
\end{tabular}  
\caption{Evolution of the QSO LF parameters versus redshift, at $0 < z < 7$. The points are their best-fit values and the relative error bars obtained through a fit by a double power-law (first row, dark red circles) and a Schechter function (second row, blue triangles). 
First row: the panels represent: (a) the logarithm of the normalization, $\log \Phi^*$, (b) the break magnitude, $M^*$, (c) the faint-end slope, $\alpha$, (d) the bright-end slope, $\beta$. Second row: (a) $\log \Phi^*$, (b) $M^*$, (c) the slope $\gamma$. In panels (a) and (b), the black curves indicate the quadratic fitting functions for the evolution of $\log \Phi^*$ and $M^*$ as a function of redshift, from $z=0$ to $z=7$, for the DPL case (solid) and the Schechter case (dashed). In panels (c) and (d) of first row, the black solid lines correspond to the mean values of $\alpha$ and $\beta$. In panel (c) of second row, the black dashed curve represents the quadratic fit for the evolution of $\gamma$. The shaded areas represent the 1$\sigma$ uncertainties on the fitted LF parameters (pink for the DPL case, cyan for the Schechter case) recovered from our Monte-Carlo sampling.}
\label{LFparameters}
\end{figure*}
We then fit these observations with two parametric luminosity function models: a double power-law function and a Schechter function. We adopt for the double power-law (hereafter referred to as DPL) the standard functional form:
\begin{align}
\Phi&(M_{\text{AB}},z)(\text{Mpc}^{-3}\text{mag}^{-1})=\nonumber\\
&\frac{\Phi^*(z)}{10^{0.4(\alpha(z)+1)(M_{\text{AB}}-M^*(z))}+10^{0.4(\beta(z)+1)(M_{\text{AB}}-M^*(z))}},
\label{DPL}
\end{align}
containing four redshift-dependent parameters: the normalization, $\Phi^*$, the break magnitude, $M^*$, the faint-end slope, $\alpha$, and the bright-end slope, $\beta$. The Schechter luminosity function is
\begin{align}
\Phi&(M_{\text{AB}},z)(\text{Mpc}^{-3}\text{mag}^{-1})=(0.4\ln 10)\Phi^*(z)\nonumber\\
&\times 10^{0.4(\gamma(z)+1)(M^*(z)-M_{\text{AB}})}\exp {\{-10^{0.4[M^*(z)-M_{\text{AB}}]}\}},
\label{SPL}
\end{align}
where the parameters to be determined are $\Phi^*$, $M^*$ and $\gamma$.

We determine the LF parameters via $\chi^2$ minimisation. The resulting best-fit values and corresponding errors (together with the $\chi^2$ value) for each redshift are listed in Table 1. In Figure \ref{lumFunc} we show our best-fit LF curves for different redshift bins in $0.5\lesssim z\lesssim 6$, along with the observational data and corresponding error bars (Jiang et al. 2009: magenta triangles; Fiore et al. 2012: purple diamond; Giallongo  et al. 2012, 2015: red circles, filled and empty respectively; Kashikawa et al. 2015: dark red square, with the arrow indicating a lower limit). Vertical green lines denote the AB magnitude of quasars that would be detectable via radio recombination lines (RRLs) (taken from M16) with the SKA-MID telescope, in 10, 100 and 1000 hours of observing time. We find that the MID instrument could detect sources with $M_{\text{AB}}\lesssim -26$ at $z<4$ and sources with $M_{\text{AB}}\lesssim -27$ at $z\gtrsim4$ in $t_{\text{obs}}<100$ hrs.

\begin{figure}
\centering
\hspace{-0.5cm}
\includegraphics[width=0.45\textwidth]{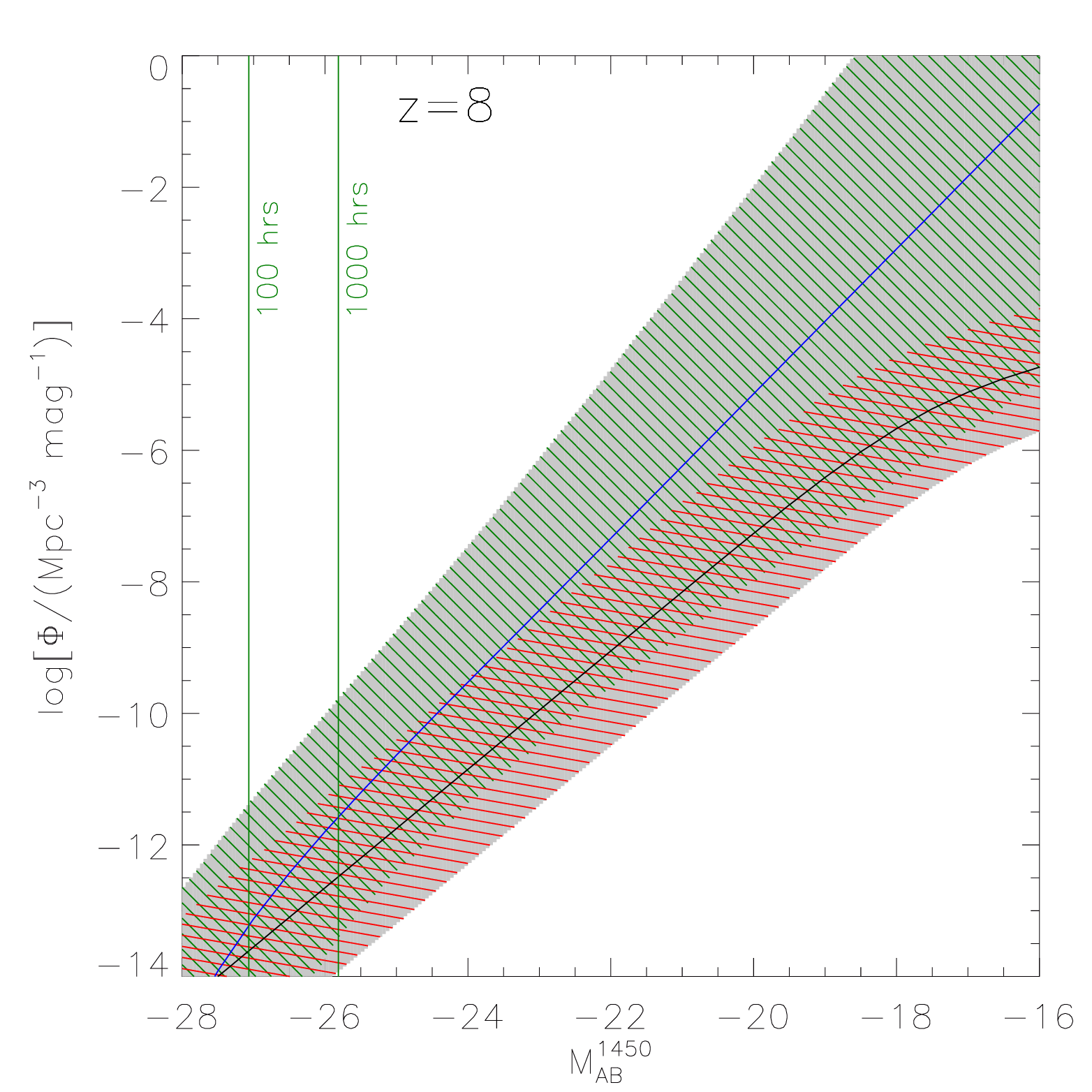} 
\caption{UV luminosity function $\log\Phi$ $(\text{Mpc}^{-3}\hspace{0.1cm}\text{mag}^{-1})$ as a function of the absolute magnitude $M_{\text{AB}}^{1450}$ at $z\sim 8$, predicted with our parametric best-fit luminosity functions, using both a double power-law (black curve, with the mean values $\alpha_{\text{mean}}\approx -1.35$, $\beta_{\text{mean}}\approx -3.23$) and a Schechter function (blue curve). 
The red and green hatched regions indicate the 1$\sigma$ uncertainties on the extrapolated LFs, in the DPL and Schechter case respectively, while the grey shaded area represents the most conservative uncertainty on the predicted LFs. We have taken into account the errors of all the parameters entering in the LF functional form using Monte-Carlo sampling. 
 The green vertical lines represent the thresholds of observability for a 5$\sigma$ detection with the SKA-MID telescope at $\nu=13.8$ GHz, in 100 and 1000 hours of integration time.
}
\label{lumFunc_z8}
\end{figure}

In order to predict the QSO LF beyond $z=6$, we derive empirical relations for $\Phi^*(z)$, $M^*(z)$, $\alpha(z)$, $\beta(z)$ and $\gamma(z)$ based on the best fit values found at lower redshifts. In Fig. \ref{LFparameters} we plot these values with the relative error bars versus redshift both in the DPL case (dark red circles in the upper panels) and in the Schechter case (blue triangles in the bottom panels). From Fig. \ref{LFparameters} we outline a few interesting features: 
\begin{itemize}
\item[(a)] $\log \Phi^*$ increases almost linearly with redshift in the DPL case, while it shows a slow increase towards $z\sim 2$ and then a significant reduction towards high redshift in the Schechter case; 
\item[(b)] $M^*$ increases until $z\sim 3$ (DPL case) or $z\sim 6$ (Schechter), and then decreases (in absolute value); 
\item[(c)] scatter in $\alpha$ and $\beta$ is large and consistent with no evolution;
\item[(d)] the evolution of $\gamma$ shows a peak at $z\sim 2.5$, followed by a gentle drop. 
\end{itemize}
The redshift evolution of $\log\Phi^*$, $M^*$ and $\gamma$ can be well fit by quadratic functions, namely: 
\begin{equation}
\begin{cases}
\log\Phi^* \approx -6.0991^{+0.1606}_{-0.1543}+0.0209^{+0.1179}_{-0.1221}z+0.0171^{+0.0196}_{-0.0190}z^2 \vspace{0.2cm}\\ M^* \approx -22.5216^{+0.2565}_{-0.2502}-1.6510^{+0.1821}_{-0.1848}z+0.2869^{+0.0280}_{-0.0280}z^2 ,
\end{cases}
\label{fit_DPL}
\end{equation}
for the DPL case, and
\begin{equation}
\begin{cases}
\log\Phi^* \approx -6.9474^{+0.1659}_{-0.1504}+0.5885^{+0.1156}_{-0.1261}z-0.1618^{+0.0201}_{-0.0186}z^2 \vspace{0.2cm}\\ M^* \approx -24.0149^{+0.2440}_{-0.2242}-0.9927^{+0.1638}_{-0.1774}z+0.0773^{+0.0271}_{-0.0251}z^2 \vspace{0.2cm}\\ \gamma \approx -2.0045^{+0.2244}_{-0.2169}+0.2330^{+0.1566}_{-0.1563}z-0.0562^{+0.0238}_{-0.0239}z^2,
\end{cases}
\label{fit_Sch}
\end{equation}
for the Schechter case. These fitting functions are estimated by performing a $\chi^{2}$ minimisation using Monte-Carlo sampling. The errors on the individual fitting coefficients correspond to the 1$\sigma$ uncertainties from the marginalised one dimensional probability distribution functions (PDFs). 
 We overplot in Fig. \ref{LFparameters} the corresponding best-fitting curves (black solid for the DPL case, black dashed for the Schechter case) with the shaded regions indicating the relative 1$\sigma$ errors (pink for the DPL case, cyan for the Schechter case).

We adopt these fitting functions to predict the quasar LF at $z>6$. In particular, Figure \ref{lumFunc_z8} presents our predictions at $z\sim 8$ both in the DPL case (namely by combining eq. \ref{DPL} and eq. \ref{fit_DPL}; black curve) and in the Schechter case (eq. \ref{SPL} and eq. \ref{fit_Sch}; blue curve). In the DPL case, we adopt the slope mean values ($\alpha = -1.35 \pm 0.91$, $\beta = -3.23 \pm 0.68$) 
 since the large uncertainties in these parameters hamper a conceivable fit to our results.

In order to estimate the errors on the extrapolated quasar LFs beyond $z>6$, we Monte-Carlo sample each of the fitting functions outlined in Eqs.~(\ref{fit_DPL}) and~(\ref{fit_Sch}), folding in the associated uncertainties on each of the individual quadratic fitting coefficients. By performing this step we are able to obtain a posterior PDF at any redshift for each of ${\rm log}\Phi^{*}$, $M^{*}$ and $\gamma$ for the Schechter case, and ${\rm log}\Phi^{*}$, $M^{*}$ for the DPL case. Finally, we perform a $\chi^{2}$ minimisation of the quasar LF by Monte-Carlo sampling the functional form (either DPL --Eq.~(\ref{DPL})-- or Schechter --Eq.~(\ref{SPL})--), assuming a joint likelihood defined by multiplying the individual posterior PDFs for the individual PDFs found above (including the errors on $\alpha$ and $\beta$ in the DPL case). Reported 1$\sigma$ errors on the extrapolated quasar LFs are then estimated from this joint likelihood distribution.

In Fig. \ref{lumFunc_z8}, the red and green hatched regions show the outcomes of our calculations for the DPL and Schechter case, respectively. As in Fig. \ref{lumFunc}, the green vertical lines indicate the thresholds of observability at $z\sim 8$ for a 5$\sigma$ detection with SKA-MID at $\nu=13.8$ GHz, in 100 and 1000 hrs of observing time. Quasars with $M_{\text{AB}}\lesssim -26$ and $-27$ can be detected with SKA-MID in $t_{\text{obs}} < 1000$ hrs and $< 100$ hrs, respectively.

We observe that the difference between the LF computed in the DPL case and the one computed in the Schechter case increases with increasing redshift. This is due to the fact that with fewer available data points for the LF, the larger the uncertainty becomes on the functional form determination. 
We actually do not know whether the DPL or a Schechter function is the correct functional form to describe the observed LF. Thus, we must consider as a conservative uncertainty on the extrapolated LFs the region between the lower limit of the DPL and the upper limit of the Schechter function (grey shaded region in Fig. \ref{lumFunc_z8}). 
\section{Implications}
\subsection{Quasar contribution to cosmic reionization}
The evolution of the quasar UV luminosity function up to high redshifts provides us important information about the AGN contribution to the ionizing UV background responsible for cosmic reionization. As already pointed out in the Introduction, the recent discovery of a considerable number of faint quasars by new deep multiwavelength AGN surveys at $z>3$ (Civano et al. 2011; Glikman et al. 2011; Fiore et al. 2012; Giallongo et al. 2015), indicates that the population of less luminous sources at high-$z$ is larger than previously estimated. This finding forces an upward revision of the ionizing power of QSOs.

With the aim of constraining the quasar contribution to reionization, we compute the redshift evolution of the AGN comoving emissivity and hydrogen photoionization rate, based on our predicted LF. First, we derive the specific ionizing emissivity following Lusso et al. (2015), which for the quasar population at redshift $z$ and frequency $\nu$ is given by
\begin{equation}
\epsilon_\nu(\nu,z)=\int_{-\infty}^{M_{\text{AB}}^{\text{min}}} \Phi(M_{\text{AB}},z)L_{\nu}(M_{\text{AB}},\nu)dM_{\text{AB}},
\end{equation}
where $L_{\nu}(M_{\text{AB}},\nu)$ is the specific luminosity as a function of $M_{\text{AB}}(\nu)$, $L_\nu=4.346\times 10^{20}\times 10^{-0.4M_{\text{AB}}(\nu)}$ erg s$^{-1}$ Hz$^{-1}$, and $M_{\text{AB}}^{\text{min}}$ is the minimum magnitude considered in the integral. We vary $M_{\text{AB}}^{\text{min}}$ from $-25$ to $-19$. 
Then, we convert the integrated emissivity at 1450 $\text{\AA}$ inferred from our UV luminosity functions into a 1 Ryd emissivity, $\epsilon_{912}$, using a power-law parametrization of the quasar SED, $f_{\nu}\propto \nu^{\alpha_{\text{EUV}}}$, with $\alpha_{\text{EUV}}=-1.7$ at $\lambda > 912$ $\text{\AA}$ (Lusso et al. 2015),
\begin{equation}
\epsilon_\nu(\nu,z)=\epsilon_{\nu,912}(z)\left (\frac{\nu}{\nu_{912}}\right )^{\alpha_{\text{EUV}}}.
\end{equation}
Finally, from the Lyman limit emissivity, we can estimate the quasar contribution to the UV background photoionization rate, by using the following relation
\begin{equation}
\Gamma_{\text{HI}}(z)=\int_{\nu_{\text{LL}}}^{\infty} \frac{4\pi J_{\nu}(\nu,z)}{h\nu}\sigma_{\text{HI}}(\nu) d\nu.
\end{equation}
Here $\nu_{\text{LL}}$ is the Lyman limit frequency, $J_{\nu}(\nu,z)$ is the mean specific intensity of the ionizing background at $z$, and $\sigma_{\text{HI}}(\nu)$ is the HI photoionization cross section\footnote{The hydrogen photoionization cross section at frequency $\nu$ is given by $\sigma_{\text{HI}}(\nu)\approx \sigma_{0}(\nu_{\text{LL}}/\nu)^3$, with $\sigma_{0}=6.33\times 10^{-18}$ cm$^2$ at 912 $\text{\AA}$.}. Using the `local-source approximation' (see Lusso et al. 2015 for details), we can rewrite the photoionization rate as follows:
\begin{align}
\Gamma_{\text{HI}}(z) &\simeq 4.6\times 10^{-13}\left (\frac{\epsilon_{\nu,912}}{10^{24}\hspace{0.1cm}\text{erg s$^{-1}$ Hz$^{-1}$ Mpc$^{-3}$}}
\right ) \nonumber\\
&\times\left (\frac{1+z}{5}\right )^{-2.4}\frac{1}{1.5-\alpha_{\text{EUV}}} \hspace{0.1cm}\text{s}^{-1}.
\end{align}

\begin{figure*}
\centering
 \begin{tabular}{@{}cc@{}}
\subfigure{\includegraphics[width=0.48\textwidth]{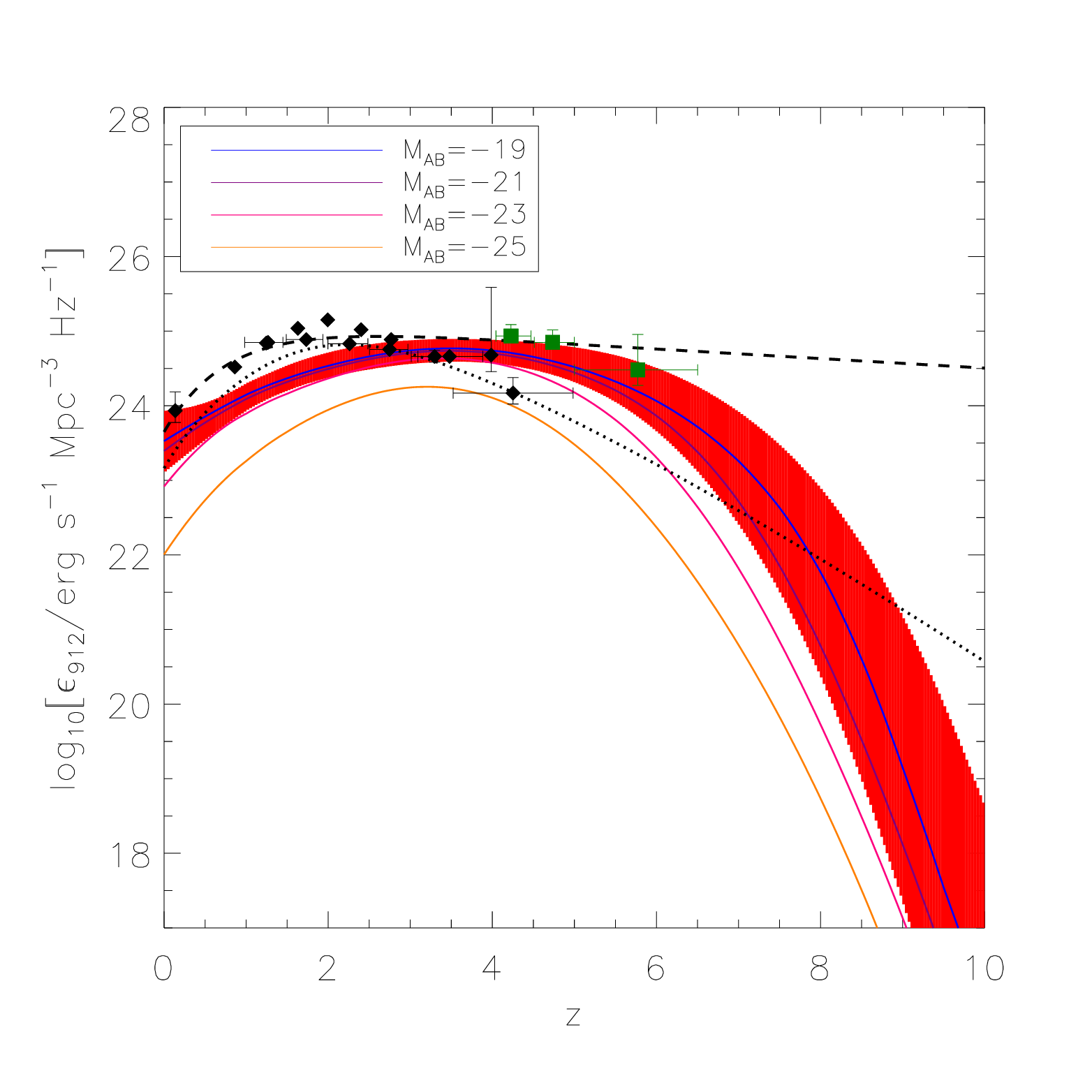}} 
       \hspace{-0.4cm}
\subfigure{\includegraphics[width=0.48\textwidth]{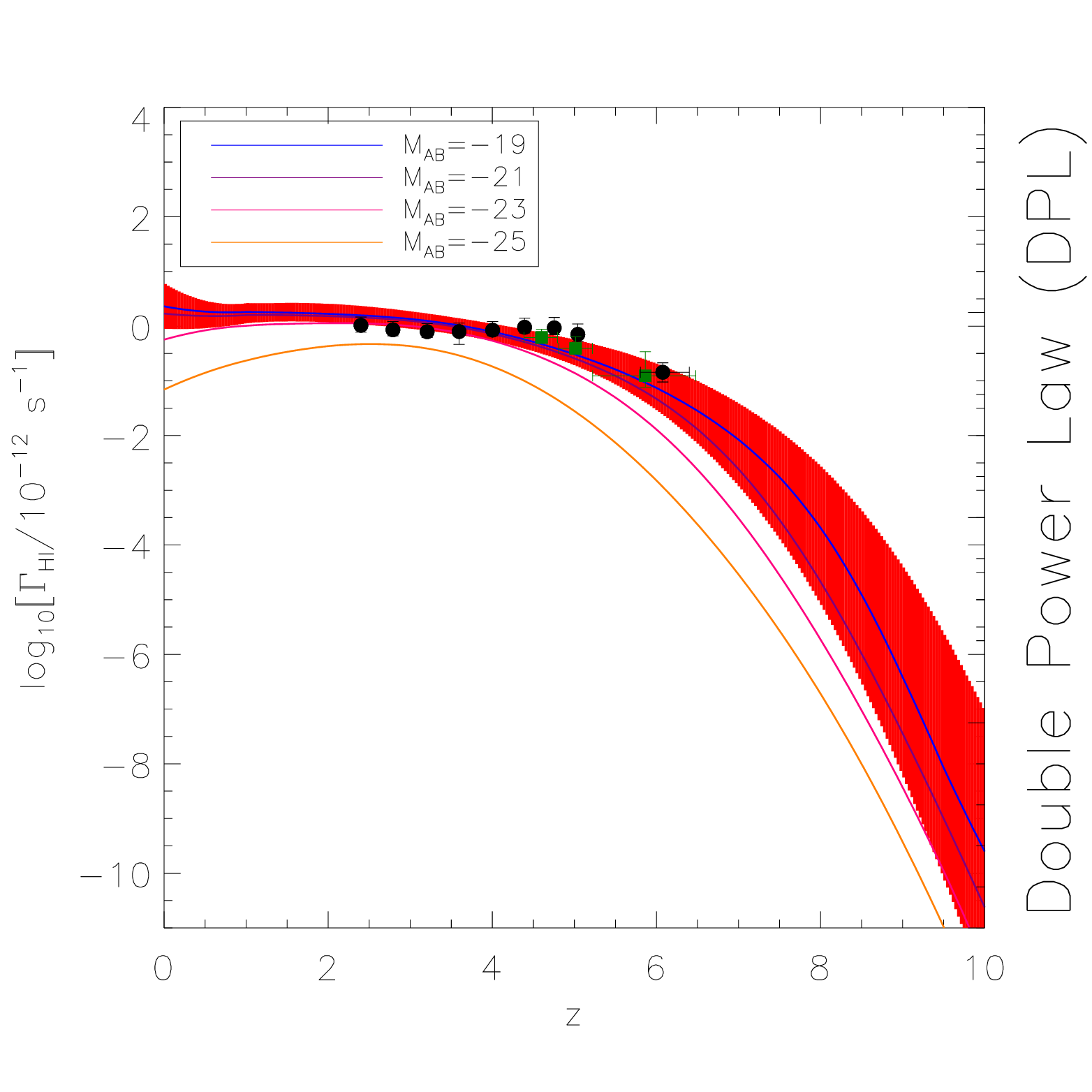}} 
\vspace{-1cm}\\
\subfigure{\includegraphics[width=0.48\textwidth]{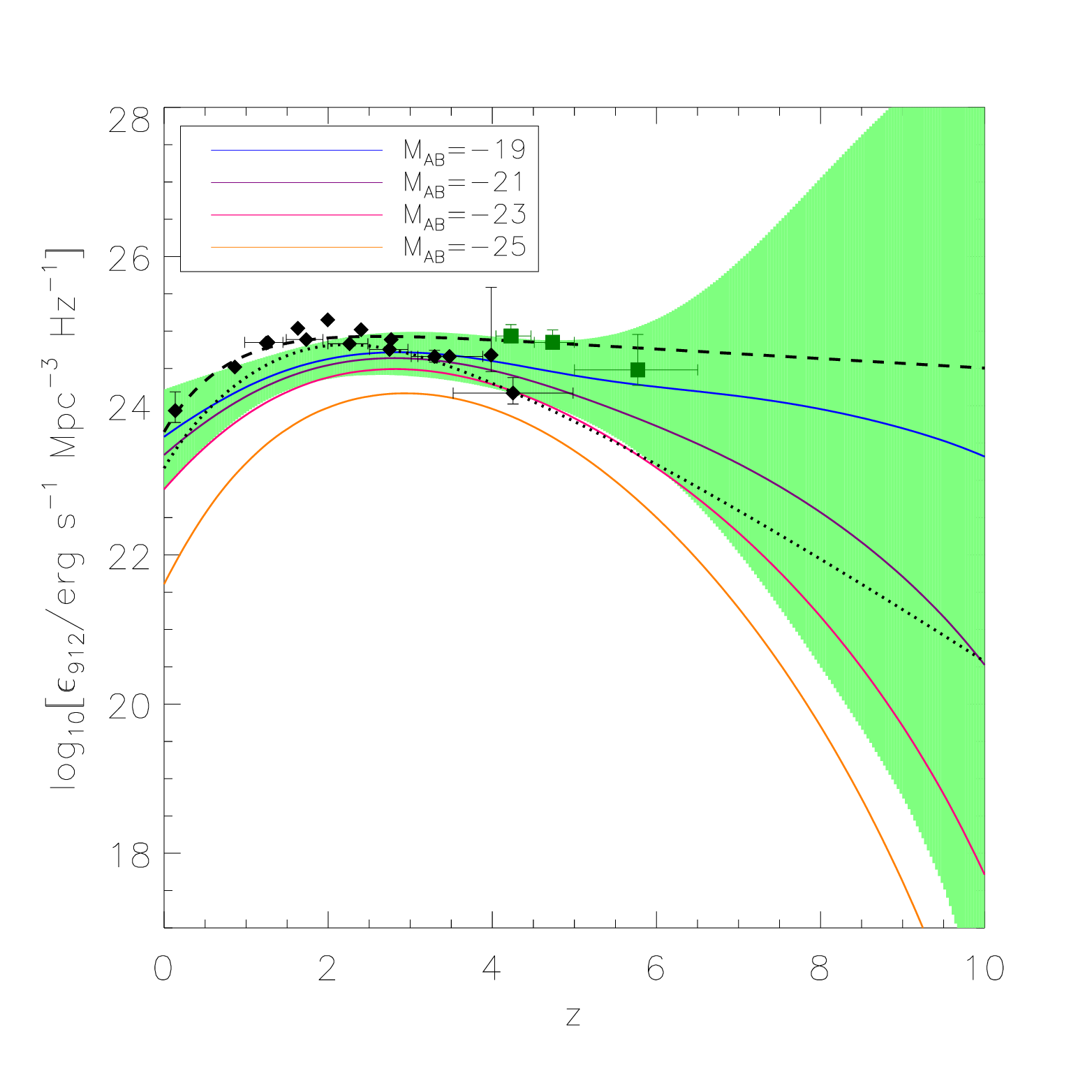}} 
        \hspace{-0.4cm}
\subfigure{\includegraphics[width=0.48\textwidth]{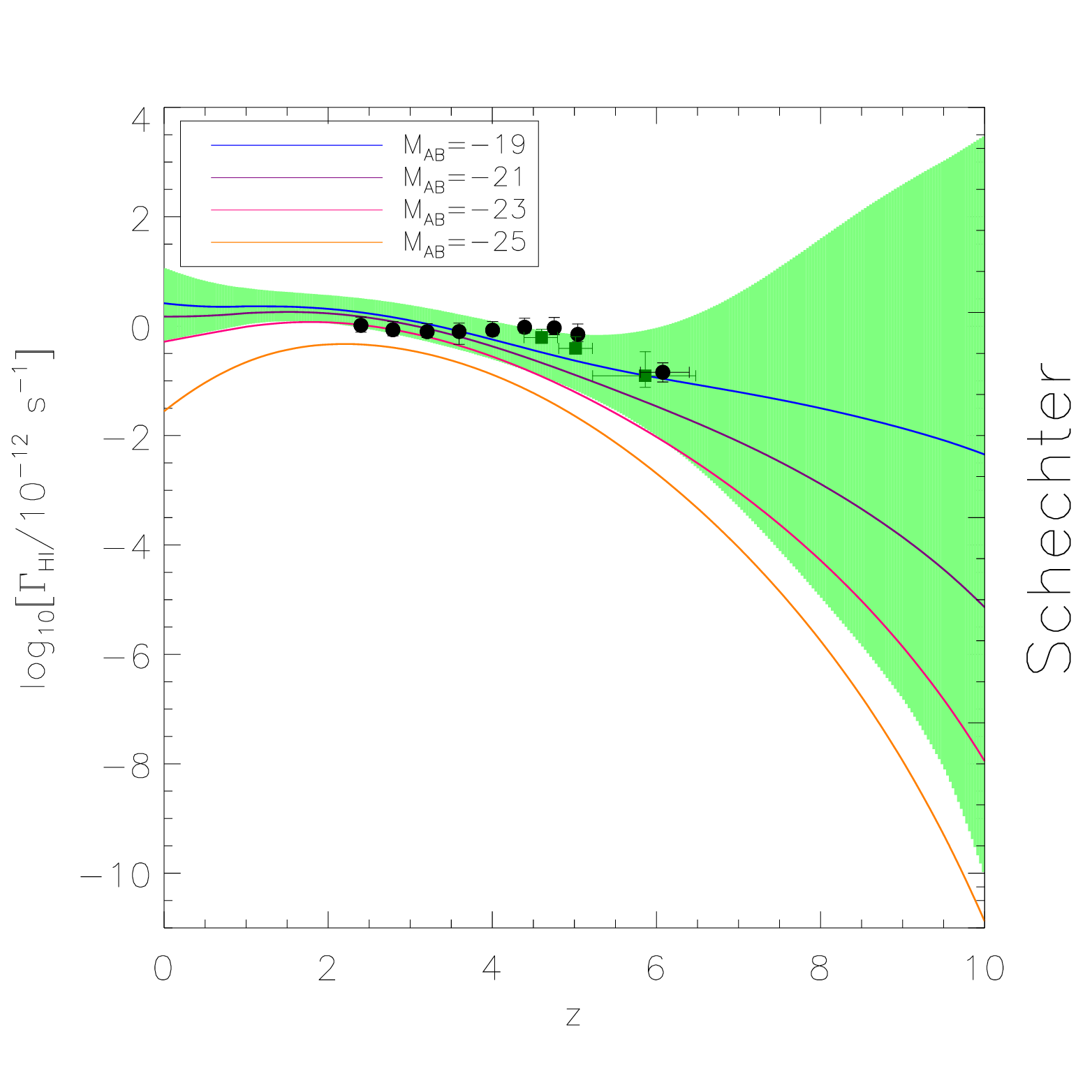}} 
\end{tabular}
\caption{Top panels: AGN comoving ionizing emissivity (left) and HI photoionization rate in units of $10^{-12}$ s$^{-1}$ (right), as a function of redshift ($0<z<10$), for a double power-law LF. Bottom: As top panels, but for a Schechter LF. In each panel, curves are inferred from our theoretical LFs using different values of the minimum absolute AB magnitude of a quasar at 1450 $\text{\AA}$ (from $M_{\text{AB}}=-25$ to $M_{\text{AB}}=-19$). The shaded regions (red in the DPL case, green in the Schechter case) indicate the uncertainty related to the variation of all LF parameters for $M_{\text{AB}}=-19$. Black diamonds in the left panels represent the measurements by Bongiorno et al. (2007); Schulze et al. (2009); Glikman et al. (2011); Masters et al. (2012); Palanque-Delabrouille et al. (2013); green squares are data from Giallongo et al. (2015). For comparison, the dotted and dashed curves show the 1 Ryd comoving emissivities from HM12 and MH15, respectively. In the right panels, black circles represent empirical measurements of the HI photoionization rate from Becker $\&$ Bolton (2013) and Calverley et al. (2011); green squares are predictions from Giallongo et al. (2015).
}
\label{emissivity_Ndot}
\end{figure*}

We present our results in Figure \ref{emissivity_Ndot}. The left (right) panels correspond to the AGN comoving emissivity (HI photoionization rate) as a function of redshift, for the double power-law (top) and Schechter (bottom) luminosity functions, respectively.
In each panel, the predicted curves are inferred from our theoretical LFs, and are plotted for different values of the minimum quasar absolute AB magnitude at 1450 $\text{\AA}$. The shaded regions (red in the DPL case, green in the Schechter case) indicate the uncertainties obtained from the Monte-Carlo sampling 
 (see Sec. 2) for $M_{\text{AB}}^{\text{min}}=-19$. In the left panels, the black diamonds represent the AGN 1 Ryd emissivity inferred from previous works (Bongiorno et al. 2007; Schulze et al. 2009; Glikman et al. 2011; Masters et al. 2012; Palanque-Delabrouille et al. 2013), while the green squares are the measurements from Giallongo et al. (2015). Moreover, we plot for comparison the curves for the quasar comoving emissivity at 1 Ryd from HM12 (dotted), which closely fits the results of H07, and from MH15 (dashed). In the right panels, the black circles are the empirical measurements of the HI photoionization rate from Calverley et al. (2011) and Becker $\&$ Bolton (2013), while the green squares represent the predicted contribution by faint AGNs from the GOODS-S sample of Giallongo et al. (2015).

Based on our results, assuming $M_{\text{AB}}^{\text{min}}=-19$, we can provide handy parameterizations for the quasar emissivity (in units of erg s$^{-1}$ Mpc$^{-3}$ Hz$^{-1}$) and photoionization rate (in units of s$^{-1}$) evolution:
\begin{equation}
\begin{cases}
\log\epsilon_{912}(z)\approx 23.59+0.55z-0.062z^2+0.0047z^3-0.0012z^4\\\log\Gamma_{\text{HI}}(z)\approx -11.66-0.081z-0.00014z^2+0.0033z^3-0.0013z^4,
\end{cases}
\label{emNdot_DPL}
\end{equation}
in the DPL case, and
\begin{equation}
\begin{cases}
\log\epsilon_{912}(z)\approx 23.55+0.99z-0.28z^2+0.031z^3-0.0013z^4\\\log\Gamma_{\text{HI}}(z)\approx -11.64+0.17z-0.13z^2+0.014z^3-0.00059z^4,
\end{cases}
\label{emNdot_Sch}
\end{equation}
in the Schechter case.

In the DPL case, the inferred emissivity peaks at a higher redshift ($z\sim 3.5$) with respect to the curve from HM12; beyond that point, it remains about 10 times larger for $M_{\text{AB}}^{\text{min}}=-19$ until $z\sim 8$, and then decreases more rapidly than the HM12 curve. In the Schechter case instead, for a minimum magnitude $M_{\text{AB}}^{\text{min}}=-19$, our LyC emissivity does not drop at high redshift (while showing a peak at $z\sim 3$ for $M_{\text{AB}}^{\text{min}}<-19$), and presents a very similar trend to the one found by MH15. For the same minimum quasar magnitude, by varying the LF parameters in the allowed intervals, we find that the emissivity (in units of $\text{erg s$^{-1}$ Mpc$^{-3}$ Hz$^{-1}$}$) at $z\sim 6$ varies in the range [$4.4\times 10^{23}$, $3.2\times 10^{24}$] for a double power-law LF, and [$1.5\times 10^{23}$, $1.5\times 10^{25}$] for a Schechter LF. This implies that in both models, our emissivity at $z\sim 6$ is higher than the one derived by H07 ($\epsilon_{912}\simeq 2\times 10^{23}$ $\text{erg s$^{-1}$ Mpc$^{-3}$ Hz$^{-1}$}$), being compatible with the value found by MH15 (see Introduction). 

As for the photoionization rate, for $M_{\text{AB}}^{\text{min}}=-19$ at $z\sim 6$ we find $\Gamma_{\text{HI}}$ (in units of $10^{-12}$ s$^{-1}$) in the range [$0.028$, $0.20$] in the DPL case, and [$0.0095$, $0.94$] in the Schechter model. These results are in agreement with the estimates from Calverley et al. (2011), $\Gamma_{\text{HI}}\simeq 0.14\times 10^{-12}$ s$^{-1}$, and Giallongo et al. (2015), $\Gamma_{\text{HI}}\simeq 0.12\times 10^{-12}$ s$^{-1}$, which are in turn consistent with the value required to keep the IGM highly ionized at $z\sim 6$. 
\begin{figure*}
\centering
 \begin{tabular}{@{}cc@{}}
 \hspace{-0.3cm}  
\subfigure{\includegraphics[width=0.35\textwidth]{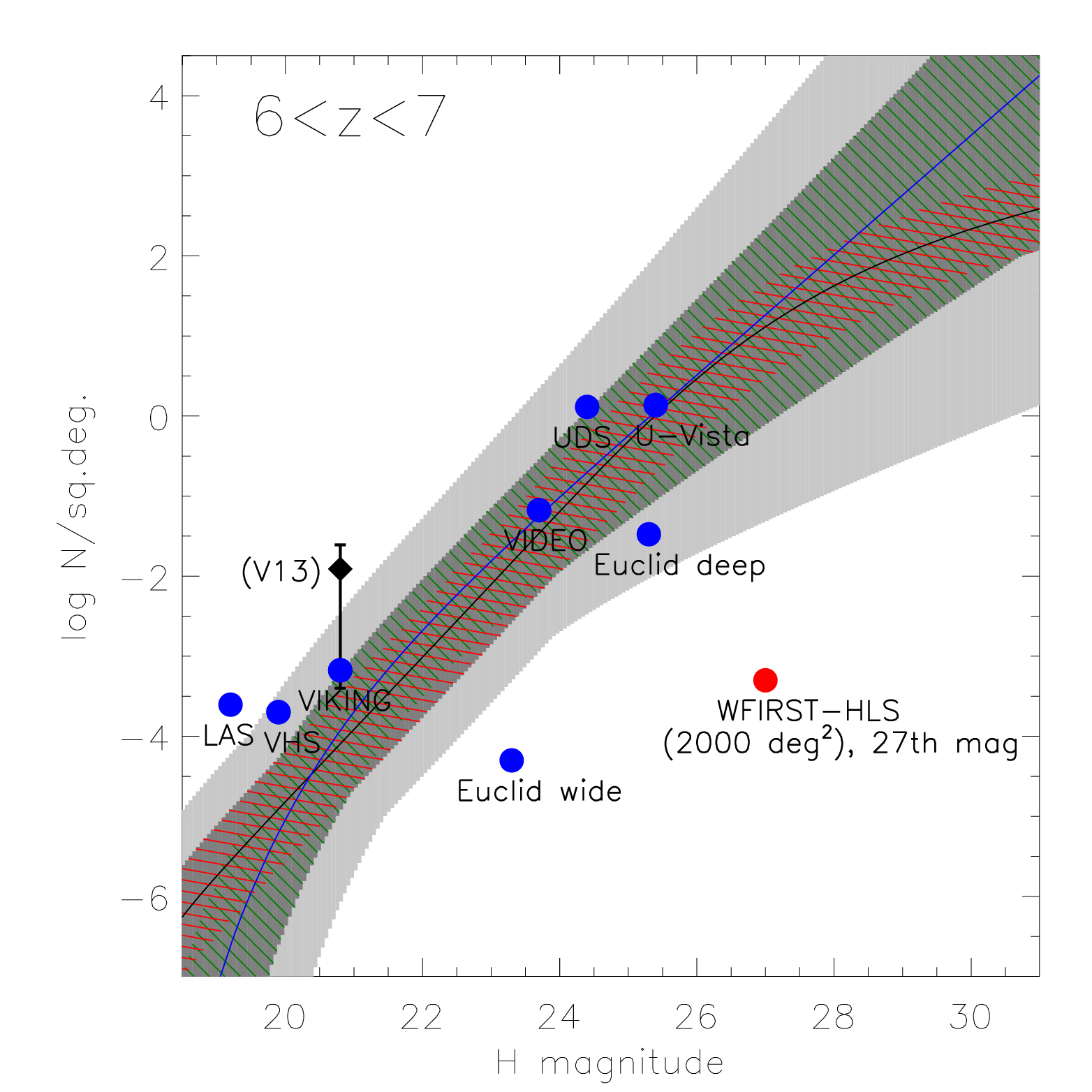}} 
        \hspace{-0.5cm}
\subfigure{\includegraphics[width=0.35\textwidth]{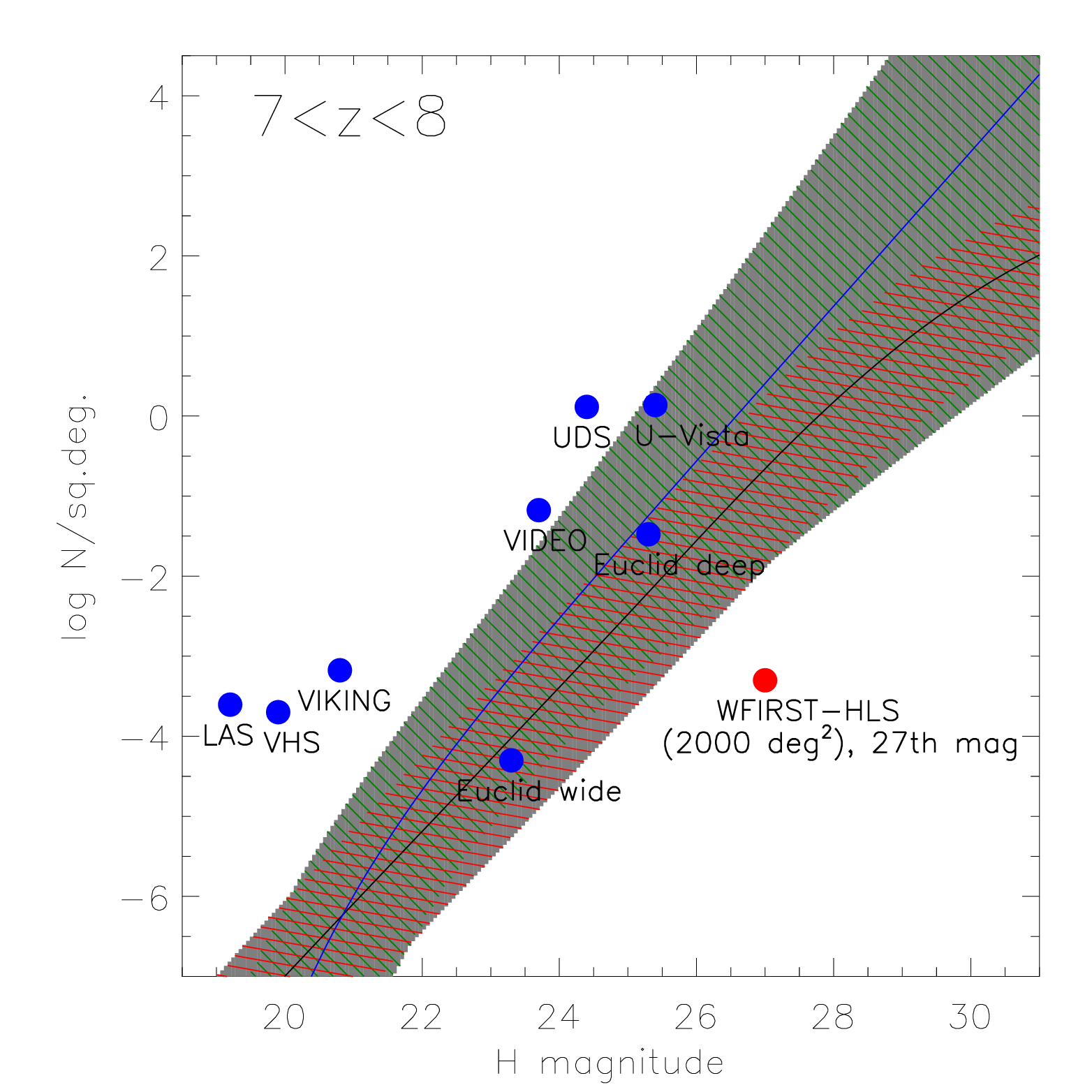}} 
        \hspace{-0.5cm} 
\subfigure{\includegraphics[width=0.35\textwidth]{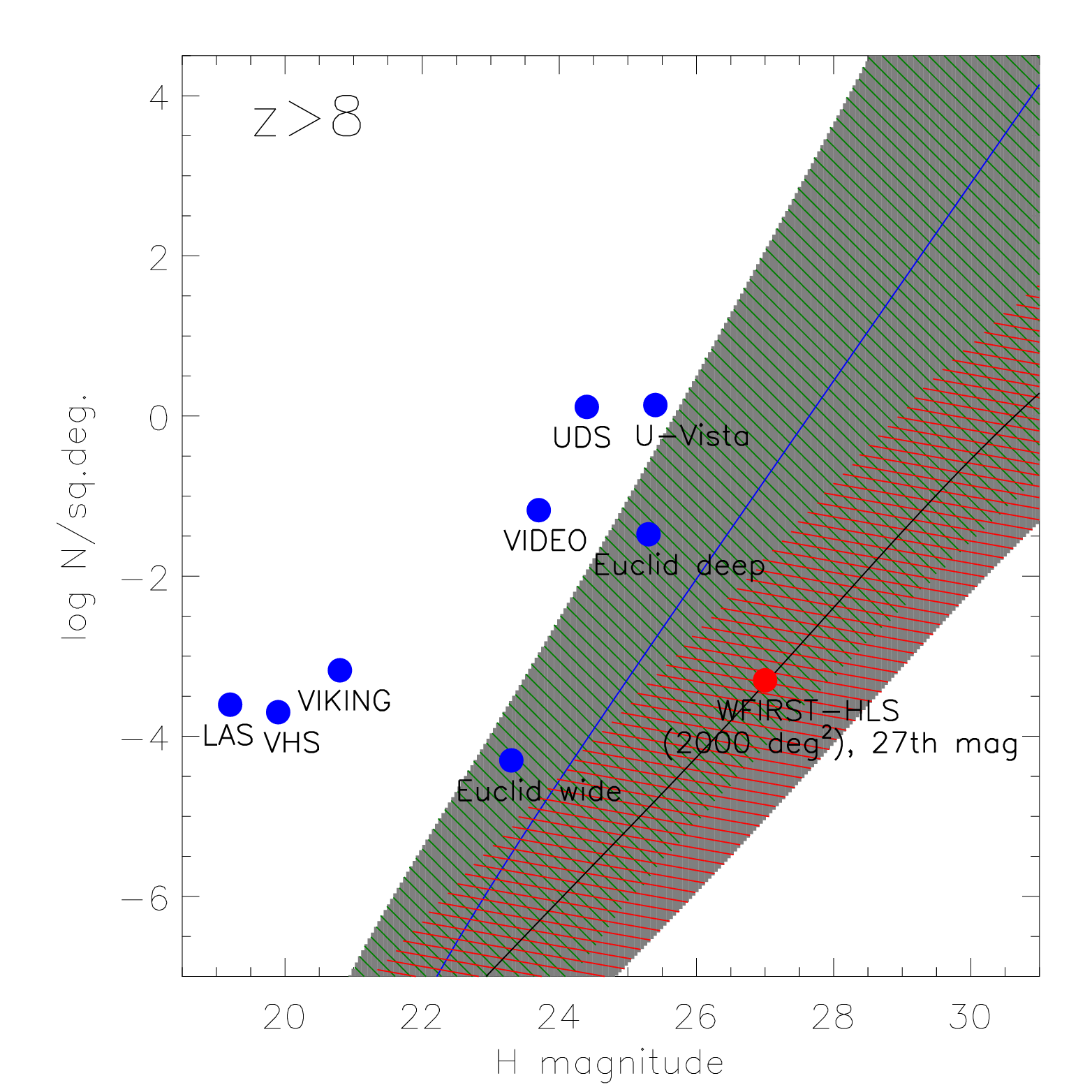}} 
\end{tabular}
\caption{Predicted sky surface density of quasars brighter than the $H$ magnitude limit, in the three redshift bins $6<z<7$, $7<z<8$, and $z>8$, using both a power-law (black curve) and a Schechter luminosity function (blue curve). The red and green hatched areas indicate the uncertainty related to luminosity functions, in the DPL and Schechter case respectively, while the grey shaded regions represent the most conservative uncertainty on the predicted number counts. The symbols show the $H$-band limiting magnitudes (at 10$\sigma$ detection) for the different surveys and the expected number density on the sky. In the redshift bin $6<z<7$, the black diamond represents the result from Venemans et al. (2013; V13) with the corresponding 2$\sigma$ error bar, while the light grey shaded area indicates the 2$\sigma$ uncertainty on the QSO sky surface density (considering both the DPL and Schechter functions).}
\label{number_countsSURVEYS}
\end{figure*}
Finally, there is a strong dependence on the minimum magnitude: varying $M_{\text{AB}}^{\text{min}}$ from $-19$ to $-25$ reduces the emissivity and photoionization rate by a factor up to $\sim 1-4$ orders of magnitude.

These results suggest that, by including the recently discovered population of faint quasars, the contribution of AGNs to the UV radiation responsible for the cosmic reionization is higher than previously found. However, we warn that such a conclusion critically depends on the adopted shape (i.e. double power-law or Schechter) for the QSO LF. 

\subsection{Near-infrared number counts}
Our study on the evolution of the QSO LF allows us to examine the possible detectability of AGNs in the NIR. In order to compute the expected quasar number counts at different redshifts, we use the following equation: 
\begin{equation}
N(<M_{\text{AB}}^{1450},z)=\int_{-\infty}^{M_{\text{AB}}^{1450}}\Phi(M_{\text{AB}},z)dM_{\text{AB}}V_c(z),
\label{Ncount}
\end{equation}
where $\Phi(M_{\text{AB}},z)$ is our parametric best-fit luminosity function (given by eq. (\ref{DPL}) or eq. (\ref{SPL}), in the DPL and Schechter case respectively), and $V_c(z)$ is the comoving volume:
\begin{equation}
V_c(z)=\frac{c}{H_0}d\Omega\int_{z}^{z+\Delta z}\frac{d_L(z')^2dz'}{(1+z')^2\sqrt{\Omega_m(1+z')^3+\Omega_\Lambda}},
\label{volume}
\end{equation}
where $c$ is the speed of light, $H_0$ is the Hubble parameter, d$\Omega$ is the solid angle corresponding to the field of view of a given telescope, $d_L$ is the luminosity distance, and $\Omega_m$ and $\Omega_\Lambda$ are the total matter density and the dark energy density in the units of critical density, respectively. Furthermore, $\Delta z$ is the redshift interval covered by the total bandwidth $\Delta\nu$ of the instrument under consideration:
\begin{equation}
\Delta z=\nu_{\text{RF}}\bigg(\frac{\Delta\nu}{\nu_{\text{obs}}^2-0.25\Delta\nu^2}\bigg),
\label{delta_z}
\end{equation}
where $\nu_{\text{obs}}=\nu_{\text{RF}}/(1+z)$ is the observed frequency ($\nu_{\text{RF}}$ is the rest-frame frequency).
 
Figure \ref{number_countsSURVEYS} shows our predicted sky surface density of high-$z$ quasars brighter than the $H$ magnitude limit, in the three redshift bins $6<z<7$, $7<z<8$, and $z>8$, obtained with either a double power-law (black curve) or a Schechter luminosity function (blue curve). The red and green hatched regions correspond to the uncertainties resulting from the Monte-Carlo sampling 
 (see Sec. 2), in the DPL and Schechter case respectively. As in Fig. \ref{lumFunc_z8}, we must consider as a conservative uncertainty on the expected number counts the region between the lower limit in the DPL and the upper limit in the Schechter case (grey shaded areas in Fig. \ref{number_countsSURVEYS}).\\
In the first panel of Fig. \ref{number_countsSURVEYS}, the black diamond represents the result from Venemans et al. (2013) with the corresponding $2\sigma$ error bar, while the light grey shaded area indicates the 2$\sigma$ uncertainty on the sky surface density in the range $6<z<7$ (considering both the DPL and Schechter functions).

As in Sec. 2, we note that the difference between our predictions with the DPL and the Schechter function increases with increasing redshift. The conversion between the absolute magnitude $M_{\text{AB}}^{1450}$ and the observed IR magnitude $H$ is done using a template quasar spectrum\footnote{We approximate quasars with a pure power-law spectrum with frequency index $\alpha_{\nu}=-0.5$.}. In the figure, the symbols show the $H$-band depth of different ongoing and future near-IR surveys and the sky density necessary for them to detect one quasar in the redshift slice. The surveys we consider here are those which could potentially discover high-$z$ quasars, i.e. cover sufficient depth and area at $H$ band to detect such kind of sources\footnote{All the $H$-band limiting magnitudes correspond to a depth of $10\sigma$.} (Willott et al. 2010). These are: 
\begin{itemize}
\item UKIRT Infrared Deep Sky Survey (UKIDSS; Lawrence et al. 2007): the Large Area Survey (LAS: 4028 deg$^2$ to $H=19.2$; Dye et al. 2006) and the Ultra Deep Survey (UDS: 0.77 deg$^2$ to $H=24.4$; Dye et al. 2006);  
\item ESO VISTA Telescope surveys (Sutherland 2009): the VISTA Hemisphere Survey (VHS: 5000 deg$^2$ to $H=19.9$), the VISTA Kilo-Degree Infrared Galaxy Survey (VIKING: 1500 deg$^2$ to $H=20.8$), the VISTA Deep Extragalactic Observations Survey (VIDEO: 15 deg$^2$ to $H=23.7$) and the UltraVISTA (U-Vista: 0.73 deg$^2$ to $H=25.4$);
\item EUCLID-wide imaging survey (20000 deg$^2$ to $H=23.3$, for weak lensing) and deep imaging survey (30 deg$^2$ to $H=25.3$, for supernovae);
\item WFIRST High Latitude Survey (WFIRST-HLS: 2000 deg$^2$ to $H=27$; Spergel et al. 2013, 2014).
\end{itemize}

The results show that in $6<z<7$, the UDS is expected to detect $\sim 1$ quasar, while the LAS may contain $\lesssim 1$ quasars (this can be considered as a lower limit; see Mortlock et al. 2012). 
Furthermore, due to the steepening of the bright end of the luminosity function at higher redshifts, the chances of detection with both UKIDSS surveys are very poor at $z>7$. Regarding the VISTA surveys, we expect to find $\sim 1$ quasar with each of them at $6<z<7$, while it is very unlikely for the four surveys to find any higher redshift sources.\\
Furthermore, from the first panel of Fig. \ref{number_countsSURVEYS}, we note that, although our predictions are below the V13 results, we are still consistent at $\sim 2\sigma$ with their observations.

The chances of detection increase significantly with EUCLID, whose wide survey should allow the discovery of $\sim 200$ quasars at $z\sim 6.5$ (in both the DPL and Schechter model), and $\sim 2$ in the DPL case ($\sim 20$ in the Schechter case) at $z\sim 7.5$. Also, EUCLID-wide is much better suited to observe distant quasars with respect to the deep survey. The capabilities of WFIRST-HLS are even more promising, since with this survey we expect to reveal $\sim 2\times 10^4$ quasars in the DPL case (up to $\sim 2\times 10^5$ in the Schechter case) at $z\sim 6.5$, $\sim 6\times 10^2$ ($\sim 6\times 10^3$) at $z\sim 7.5$, and $\sim 1$ ($\sim 200$) at $z\sim 8.5$.

We finally note, that surveys of high-$z$ QSOs are always plagued by the possible presence of contaminants (e.g. brown dwarfs). Thus, objects detected down to the survey magnitude limit may not always be unequivocally identified as actual QSOs.
\subsection{Radio number counts} 
By combining eqs. (\ref{Ncount}), (\ref{volume}) and (\ref{delta_z}), we can also predict the expected quasar number counts for a single observation with SKA-MID (FoV = 0.49 $\text{deg}^2$). Here, based on M16, we analyse the detectability of obscured AGNs in the radio band through their RRL emission\footnote{Radio recombination lines represent a special class of spectral lines arising in HII regions from transitions between highly excited hydrogen levels (quantum numbers $n>27$) and appearing in the radio regime (rest frame frequencies $\nu_e<300$ GHz). For a comprehensive review on RRLs, see Gordon \& Sorochenko (2002).}. Specifically, we consider the H$n\alpha$ lines, which are the strongest RRLs, i.e. due to $n+1 \rightarrow n$ transitions, whose rest frequencies are:
\begin{equation}
\nu_{\text{RF}}(z,n)=c\hspace{0.05cm}\text{R}_{\text{H}}\left[\frac{1}{n^2}-\frac{1}{(n+1)^2}\right],
\label{nu_obs}
\end{equation}
where $\text{R}_{\text{H}}=1.0968 \times 10^5$ $\text{cm}^{-1}$ is the Rydberg number for hydrogen, and $n$ is the principal quantum number. We compute the RRL fluxes by adopting our ``fiducial'' case for the AGN spectral indices\footnote{We parametrize the emitted quasar flux per unit wavelength as $f_{\lambda} \propto \lambda^{\alpha}$. The ``fiducial'' model is based on the following combination of spectral slopes: $\alpha_{\text{X,hard}}=-1.11, \alpha_{\text{X,soft}}=-0.7, \alpha_{\text{EUV}}=-0.7, \alpha_{\text{UV}}=-1.7$.}, with the inclusion of secondary ionizations from X-ray photons (assuming a fully neutral medium surrounding the HII region; for details see M16):
\begin{align}
f_{\text{H}n\alpha} &\approx 4.78\times 10^{-7}n^{-2.72}10^{-0.4M_{\text{AB}}^{1500}}(1+z) \nonumber\\
&\times \left(\frac{d_L}{10^5 \hspace{0.1cm}\text{Mpc}}\right)^{-2} \left(\frac{\delta v}{100 \hspace{0.1cm}\text{km s$^{-1}$}}\right)^{-1}(1-f_{\text{esc}}^{912}) \hspace{0.1cm}\mu \text{Jy}.
\label{fHnalpha}
\end{align}
Here $d_L$ is the luminosity distance of the emitting source at redshift $z$, $\delta v$ is the width of the line in velocity units, and $f_{\text{esc}}^{912}$ is the escape fraction of ionizing photons\footnote{We assume $f_{\text{esc}}^{912}=0$, i.e. all the ionizing photons remain trapped within the interstellar medium of the quasar host galaxy, and $\delta v=100$ km s$^{-1}$.}.

\begin{figure}
\centering
\hspace{-0.5cm}
\includegraphics[width=0.5\textwidth]{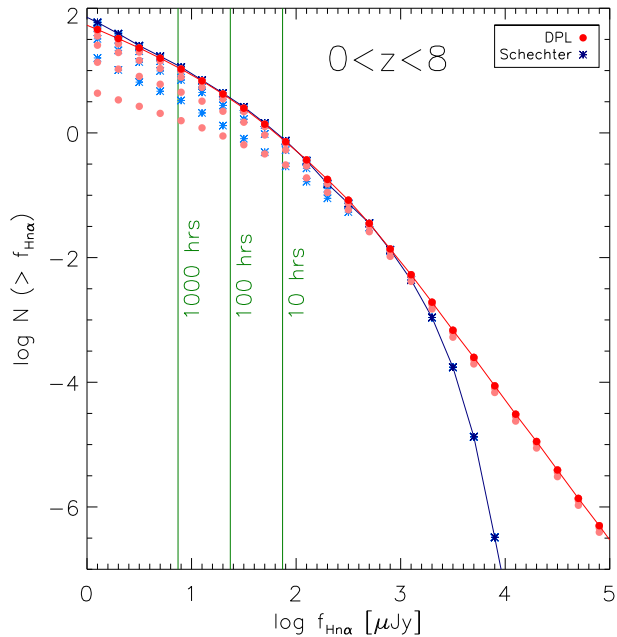}
\caption{Logarithm of quasar number counts $\log\text{N}(>f_{\text{H}n\alpha})$ as a function of the logarithm of the RRL flux density $\log f_{\text{H}n\alpha}$ 
(in $\mu$Jy), predicted for a single observation with SKA-MID at $\nu=13.5$ GHz, at $0<z<8$. The calculations are done assuming both a double power-law (pink and red circles) and a Schechter luminosity function (cyan and blue stars). The pink circles (cyan stars) refer to curves corresponding to the cumulative contribution of each redshift slice to the quasar number counts, in such a way that the red circles (blue stars), connected with a solid line, give the total expected number of sources at $0<z<8$. The green vertical lines indicate the thresholds of observability for a 5$\sigma$ detection with SKA-MID at $\nu=13.5$ GHz, in 10, 100 and 1000 hours of integration time.
}
\label{number_counts_fHnalpha}
\end{figure}

In Fig. \ref{number_counts_fHnalpha} we plot our prediction for the number counts $\log N(>f_{\text{H}n\alpha})$ of quasars contained in the volume $V_c$ spanned by SKA-MID in one pointing, at $\nu_{\text{obs}}=13.5$ GHz, as a function of the flux density at the center of H$n\alpha$ lines (in $\mu$Jy), at $0<z<8$. In our calculations we assume an instantaneous bandwidth of SKA-MID $\Delta\nu=0.5$ GHz. The computation is done by splitting the redshift interval into eight $\Delta z =1$ slices in which a certain number of radio recombination lines fall in the SKA-MID bandwidth (with $n_{\text{min}}<n<n_{\text{max}}$). 

Given that $f_{\text{H}n\alpha}=f(M_{\text{AB}}^{1450},z,n)$, for every $z$-bin and for every quasar magnitude $M_{\text{AB}}^{1450}$, a value of the H$n\alpha$ flux density corresponds to each observable quantum number $n$, and the total RRL intensity is given by the sum of the peak fluxes of all the lines,
\begin{equation}
f^{\text{tot}}_{\text{H}n\alpha}(M_{\text{AB}}^{1450},z)=\sum_{n=n_{\text{min}}}^{n_{\text{max}}}f_{\text{H}n\alpha}(M_{\text{AB}}^{1450},z,n). 
\end{equation}
Hence, in  each $\Delta z$, a given magnitude interval spans a range of $f_{\text{H}n\alpha}^{\text{tot}}$ values, and the corresponding number counts are $N_{\Delta z}(>f_{\text{H}n\alpha}^{\text{tot}})$. Thus, after the rebinning of the $f_{\text{H}n\alpha}^{\text{tot}}$ range, the total expected number of quasars at $0<z<8$ is given by the sum of the number counts in each redshift interval, $N(>f_{\text{H}n\alpha}^{\text{tot}})=\sum_{\Delta z}N_{\Delta z}(>f_{\text{H}n\alpha}^{\text{tot}})$.

The results are shown in Fig. \ref{number_counts_fHnalpha} for our predicted luminosity functions, using both a double power-law (red circles) and a Schechter function (blue stars). 
The pink circles for the DPL case (cyan stars for the Schechter case) show the cumulative contribution of each redshift slice to the quasar number counts, in such a way that the red circles (blue stars), connected with a solid line, give the total expected number of sources in $0<z<8$. The green vertical lines indicate the threshold of observability for a 5$\sigma$ detection with SKA-MID at $\nu=13.5$ GHz, in 10, 100 and 1000 hrs of integration time. Intriguingly, the MID telescope could detect RRLs with $f_{\text{H}n\alpha}\gtrsim (75, 25, 7)$ $\mu$Jy in $t_{\text{obs}}< (10, 100, 1000)$ hrs, respectively. 

\section{Summary}
We have studied the redshift evolution of the quasar UV LF at $0.5<z<6.5$, taking into account the most updated observational data and, in particular, the new population of faint quasars recently discovered (Giallongo et al. 2015).

We have fitted the quasar LF at different redshifts both with a double power-law (DPL) and a Schechter function, finding that both forms provide good fits to the data. Handy and accurate analytical fitting formulae for the redshift evolution of the normalization parameter ($\Phi^*$), the break magnitude ($M^*$) and the faint-end/bright-end slopes ($\alpha$ and $\beta$) in the double power-law case and the Schechter slope ($\gamma$) are provided.  

We first examined the implications of our results to quantify the ionizing contribution of QSOs during the EoR. We have found that, by including the new faint population of quasars, the QSO/AGN contribution is higher than previously determined, consistently with Madau \& Haardt (2015), but the level of contribution depends sensitively on the LF shape (i.e. DPL or Schechter). Notably, the LyC emissivity in the Schechter case does not drop at $z\gtrsim 3.5$ like in the DPL case, resulting in an upper limit for $\epsilon_{912}$ at $z\sim 6$ about $4$ times larger than in the DPL case. Both models shake the traditional view of an evolution of the ionizing QSO background peaking at $z=2-3$ and then quickly decreasing\footnote{While the paper was under the process of refereeing, Jiang et al. (2016) have published a study in which, based on the final SDSS high-redshift quasar sample, they have derived the $z\sim 6$ QSO luminosity function and constrained the fitted LF parameters. Using this fitted LF, they have estimated the QSO contribution to the ionizing background at $z\sim 6$. According to their results, the observed QSO population is not sufficient to ionize the IGM at $z\sim 6$. The difference between their and our analysis is that in our work we have taken into account the recently discovered population of $z>4$ low-luminosity AGNs by Giallongo et al. (2015). The main aim of this paper is in fact to examine the impact of these faint quasars on the contribution of AGNs to cosmic reionization and on their expected number counts at $z>6$.}. We have also derived an estimate for the hydrogen photoionization rate which, in both cases, is consistent with the most up to date estimates at $z\sim 6$. Given the strong dependence of these results on the shape of the UV LF, further observations of quasars at high redshift are required.

To this aim, we have made predictions for the number of $z>6$ quasars that may be discovered by current and future NIR surveys. While UKIDSS and VISTA may find a few sources at $6<z<7$, the chances of observation increase dramatically for EUCLID, which will  be able to reveal hundreds of quasars at $6<z<7$ and tens of quasars at $7<z<8$. Finally, detections of up to 200 sources at $z=8.5$ are expected with the WFIRST-HLS survey.

As a complementary strategy, we have computed the expected QSO number counts for a single radio-recombination line observation with SKA-MID as a function of the H$n\alpha$ flux density, at $0<z<8$. Intriguingly, the MID telescope could detect RRLs with $f_{\text{H}n\alpha}\gtrsim (75, 25, 7)$ $\mu$Jy in $t_{\text{obs}}< (10, 100, 1000)$ hrs, respectively.

\end{document}